\documentclass[preprint,showpacs,preprintnumbers,superscriptaddress]{revtex4-1}
\usepackage{amsmath,amssymb,graphicx,bm,subfigure,float,siunitx,color}
\usepackage{natbib}
\newcommand{\be}{\begin{equation}}
\newcommand{\ee}{\end{equation}}

\newcommand{\1}{\left}
\newcommand{\2}{\right}
\def\({\left(}
\def\){\right)}
\def\[{\left[}
\def\]{\right]}

\newcommand{\dif}{\,\mathrm{d}}

\newcommand{\m}{\mu}
\newcommand{\n}{\nu}
\newcommand{\al}{\alpha}

\newcommand{\na}{\nabla}

\begin{document}

\title{\boldmath The stability of the de-Sitter universe in nonlocal gravity}

\author{Haiyuan Feng\footnote{Corresponding author}}
\email{Email address: fenghaiyuanphysics@gmail.com }
\affiliation{School of Physics and Electronic Engineering, Shanxi Normal University, Taiyuan 030031, China}

\author{Rong-Jia Yang\footnote{Corresponding author}}
\email{Email address: yangrongjia@tsinghua.org.cn}
\affiliation{College of Physical Science and Technology, Hebei University, Baoding 071002, China}

\begin{abstract}
We constructed the ghost-free condition for nonlocal gravity using de-Sitter background field expansion and identified the structure of the nontrivial form factors. Our analysis shows that the particle spectrum of this model is nearly equivalent to general relativity (GR), with the potential addition of a scalar particle with positive mass $m$. Additionally, by employing recursion relations, we established the equivalence between nonlocal gravity and higher-derivative gravity. Moreover, we provided a comprehensive  proof of the stability of de-Sitter solution within the nonlocal framework.
\end{abstract}

\maketitle

\section{Introduction}
Recent cosmological observations, including those coming from Supernovae Ia (SNe Ia), the cosmic microwave background (CMB) radiation, large scale structure (LSS), baryon acoustic oscillations (BAO), and weak lensing, provide the means to impose combined constraints on cosmological parameters \cite{Perlmutter_1999,Riess_1998,Spergel_2003,Spergel_2007,Komatsu_2009,Komatsu_2011,Eisenstein_2005,Jain_2003,Kilbinger:2008gk}. These observations consistently suggest that the universe is currently undergoing an accelerated expansion. The observed acceleration is generally attributed to an effective positive cosmological constant, which is linked to the dark energy problem. Dark energy, which accounts for this accelerated expansion, is strongly supported by numerous astronomical observations \cite{SupernovaCosmologyProject:1998vns, SupernovaSearchTeam:1998fmf, Astier_2006, SDSS:2003eyi, WMAP:2010qai, SDSS:2004kqt, SDSS:2005xqv, Jain:2003tba}. Specifically, the dark energy problem has been well predicted within the framework of the Standard Model of Cosmology ($\Lambda$CDM), which is based on General Relativity (GR). GR has demonstrated remarkable success in the infrared (IR) regime, accurately predicting and aligning with a wide array of empirical observations, including tests within the solar system and broader cosmological phenomena. Despite its achievements, GR faces significant challenges in the ultraviolet (UV) regime, where it remains incomplete both classically and quantum mechanically. The theory encounters singularities in black holes and cosmology, with quantum corrections leading to non-renormalizability beyond the one-loop level. While black hole singularities are covered by event horizons, cosmological singularities remain exposed, causing energy densities and curvatures to diverge as physical time approaches zero. \cite{t1974one,Deser:1974cz,Deser:1974xq,Hawking:1973uf,tHooft:1974toh}.  Additionally, the equation of state (EoS) parameter for dark energy predicted by GR is $w = -1$ (with $p = w \rho$). If this value were precise, it would confirm GR with a cosmological constant. However, current astronomical data do not entirely rule out small deviations from this value. Furthermore, neither the sign nor the trend of such deviations is definitively known at present.

These issues permit a variety of theoretical models, each based on distinct fundamental theories, to address this ambiguity. One notable avenue of exploration in this field is quantum $R^2$ gravity \cite{Buchbinder:1992rb}, which effectively addresses early cosmic inflation by incorporating additional curvature terms. However, due to its inadequacies in the UV regime, it was ultimately discarded. Subsequently, the straightforward $f(R)$ model, which extends directly to the curvature scalar $R$, has been examined for its quantum behavior \cite{Codello:2007bd,Machado:2007ea,Codello:2008vh,Knorr:2019atm}. This model offers insights into both the inflationary phase of the early universe and the subsequent late-time accelerated expansion. Initial one-loop divergent calculations for $f(R)$ gravity in maximally symmetric spacetime were reported \cite{Cognola:2005de}, and these results have since been extended to more general scenarios \cite{Ruf:2017bqx}. Due to its toy-model nature, it is difficult to develop a unified framework that addresses all of these issues comprehensively. Additionally, higher-order gravity \cite{Alvarez-Gaume:2015rwa}, which incorporates contributions from higher-order curvature tensors \cite{Salam:1978fd,Masuda:1976qg,Tomboulis:1983sw}, represents another approach to modifying gravity. Nevertheless, this model encounters challenges, such as the introduction of ghost particles with spin-2 mass, which exhibit non-unitary behavior in its original quantization according to the Feynman prescription \cite{PhysRevD.16.953}. Consequently, various promising approaches have been explored to address the unitarity issue \cite{Tomboulis:1977jk, Tomboulis:1980bs, Antoniadis:1986tu}.

One of the most promising theories in the realm of modified gravity is nonlocal gravity \cite{Modesto:2017sdr, Belgacem:2017cqo, Koshelev:2016xqb}. The model modifies the Newtonian potential, smoothing out the singularity at the origin. Specifically, this potential exhibits a universal behavior, approaching a constant limit at zero distance for a broad class of nonlocal functions, while naturally recovering the standard $\frac{1}{r}$ falloff at large distance. Additionally, various cosmologically relevant bounce solutions have been constructed and extensively analyzed. At the perturbative level, the model successfully accommodate inflationary scenarios, including Starobinsky inflation \cite{Koshelev:2016xqb,Starobinsky:1980te,Koshelev:2017tvv}. 
At the quantum level, the nonlocal gravity is shown to be renormalizable through power-counting technique, with unitarity preserved. This indicates that there are explicitly defined conditions on the nonlocal functions of the d'Alembert operator $\Box$ that ensure a ghost-free spectrum of physical excitations while maintaining renormalizability during quantization \cite{Talaganis_2015,Don__2015}.

In this paper, we examine the nonlocal gravity model with a particular emphasis on the stability of the de-Sitter solution. The stability of such solution is crucial in various theoretical frameworks. For example, in the $\Lambda$CDM model, ensuring stability is essential to prevent future singularity. Alternatively, the unresolved cosmological constant problem complicates the situation. In contrast, modified gravity models, as previously discussed, offer a potentially natural geometric perspective that is consistent with Einstein's original ideas. Thus, understanding the stability or instability of de-Sitter solutions within these modified gravity models is interest. Specifically, the stability issues of nonlocal gravity have been well-explored \cite{Calcagni:2017sov,Briscese:2019rii,Briscese:2018bny,Calcagni_2018}. Nevertheless, these solutions have predominantly been addressed from a perturbative technique. Our investigation aims to provide a new perspective to investigate the stability of de-Sitter solution.

The paper is organized as follows: In Section II,  we will present a comprehensive review of super-renormalizable nonlocal gravity, focusing specifically on modifications of the forms $RF_{0}(\Box)R$ and $R_{\m\n}F_2(\Box)R^{\mu\nu}$. In Section III, we will examine the particle spectrum within de-Sitter background and demonstrate that the nonlocal gravity is nearly equivalent to GR, with the notable exception of an additional excitation of a scalar particle with positive mass. We will also establish the ghost-free condition and analyze the model's stability using the eigenvalue properties of the Laplace operator. In Section IV, we will establish the equivalence between nonlocal gravity and higher-derivative gravity using recursion relations and outline the conditions under which this equivalence is valid. By examining the stability concerns associated with higher-derivative gravity, we will provide a robust demonstration of the stability of the de-Sitter solution within the nonlocal gravity framework. The final Section will summarize our main conclusions.

\section{The string-inspired Nonlocal gravity }
Due to the inherent limitations of GR, a variety of modified gravity models have naturally emerged \cite{Nojiri:2017ncd,Nojiri:2010wj,Nojiri:2006ri,Capozziello:2011et,Faraoni:2010pgm,delaCruz-Dombriz:2012bni,Olmo:2011uz}.  Among these, string-inspired nonlocal gravity models have attracted significant attention and support, becoming a focal point of theoretical research due to their more favorable quantum behavior \cite{Deser:2007jk}. Initially, some nonlocal theories were proposed to explain the accelerated expansion of the universe and later evolved into frameworks for describing quantum phenomena \cite{NOJIRI2008821, Bamba:2012ky, Zhang:2011uv, Elizalde:2011su, Nojiri:2010pw, Zhang:2016ykx}. The profound impact of these models is evidenced by the incorporation of nonlocal interaction terms, which are also present in string theory \cite{Nojiri:2007uq, Jhingan:2008ym}. To describe physical phenomena, most nonlocal quantum gravity models introduce either nonlocal scalar fields or the d'Alembertian operator $\Box$. Without loss of generality, we focus on a general nonlocal model, represented by the following action
\be
\label{1}
S=\int \dif^4x\sqrt{-g}\[\frac{M^2_P}{2}R+\frac{\lambda}{2}\(RF_{0}(\Box)R+R_{\m\n}F_2(\Box)R^{\mu\nu}+R_{\m\n\sigma\rho}F_4(\Box)R^{\m\n\sigma\rho}\)-\Lambda+V  \],
\ee
where $R$ is the Riemann curvature scalar, $\Lambda$ is the cosmological constant, and a set of local terms $V$ cubic or higher in curvature. $M_P$ is the Planckian mass and $\lambda$ is a dimensionless parameter measuring the effect of the $O(R^2)$ corrections. The crucial elements of our analysis are the functions of the covariant d'Alembertian operator $F_i(\Box)$ which called form factors. These form factors are assumed to be entire functions, allowing them to be expanded in a Taylor series $F_{i}(\Box)=\sum^{\infty}_{n=0}f_{in}\Box^n/M^{2n}_{\ast}$. ($M_{\ast}$ represents the mass scale at which the higher derivative terms in the action gain significance.) Specifically, the model can be viewed as an infinite-order derivative extension of higher derivative gravity. Higher derivative model exhibit massive ghost particle excitations at the Planck scale $M_{P}$ and drive inflation through scalar particle excitations at the scale $m$ \cite{PhysRevLett.116.031302,PhysRevD.16.953,Koshelev_2016,Planck:2015fie,Planck:2015sxf}. Therefore, the range of $M_{\ast}$ can naturally be chosen as $m<M_{\ast}<M_{P}$.

Additionally, the super-renormalization and unitarity constrain the form factors to the following types \cite{Koshelev_2018}
\be
\1\{\begin{split}
\label{2}
&F_{0}(\Box)=-\frac{M^2_p}{\lambda}\(\frac{2\(e^{H_{0}(\Box)}-1\)+4\(e^{H_{2}(\Box)}-1\)}{12\Box}\)+F_{4}(\Box),\\
&F_{2}(\Box)=\frac{M^2_p}{\lambda}\(\frac{e^{H_{2}(\Box)}-1}{\Box}\)-4F_{4}(\Box),
\end{split}\2.
\ee
where $F_{4}(\Box)$ remains arbitrary, super-renormalizability necessitates that it shares the same asymptotic UV behavior as the other two form factors $F_{i}(\Box)$ ($i=0,2$). To achieve this configuration, the minimal approach involves retaining only two of the three form factors, which allows us to set $F_{4}(\Box)=0$. The entire function $e^{H_{i}(\Box)}$ must satisfy the three categories of conditions \cite{tomboulis1997superrenormalizablegaugegravitationaltheories}:

$\bullet$ The function $ e^{H_{i}(\Box)}$ must be real and positive along the real axis and have no zeros within the entire complex plane for $ z < \infty$ $\(z\equiv-\frac{\Box}{M^2_{\ast}}\)$. This requirement guarantees the absence of gauge-invariant poles, except for the transverse massless physical graviton pole.

$\bullet$ $e^{H_{i}(\Box)}$ exhibits the same asymptotic behavior along the real axis at $\pm\infty$.

$\bullet$ There exist value $ 0<\Phi <\frac{\pi}{2} $, and a positive integer $\gamma $, such that asymptotically
\be
\label{3}
|e^{H_{i}(\Box)}|\rightarrow|z|^{\gamma+1},\quad |z|\rightarrow\infty,\quad \gamma\geqslant2,
\ee
with regin $C$
\be
\label{4}
C\equiv\{z|-\Phi<argz<+\Phi,\pi-\Phi<argz<\pi+\Phi\}.
\ee

This final condition is crucial for ensuring optimal convergence of the theory in the UV regime. The required asymptotic behavior must be enforced not only along the real axis but also within the surrounding region $C$. Ref. \cite{tomboulis1997superrenormalizablegaugegravitationaltheories} provides an example that satisfies these three conditions, where the form factors can be expressed as
\be
\label{5}
e^{H_{i}(\Box)}=e^{\frac{1}{2}\[\Gamma(0,p_{i}(z)^2)+\gamma_{E}+\log\(p_{i}(z)^2\) \]},
\ee
where $\gamma_E \approx 0.577216$ denotes the Euler-Mascheroni constant, $\Gamma(0, z) = \int_{z}^{\infty} \frac{e^{-t}}{t} \dif t$ represents the incomplete Gamma function with its first argument set to zero. The polynomial  $p_i(z)$, which has a degree $\gamma + 1$ and satisfying $p_i(0) = 0$, ensures that the low-energy limit of nonlocal theory is correct. In the UV regime ($|z| \gg 1$), the function exhibits polynomial behavior $|z|^{\gamma + 1}$ in conical region around the real axis, with an angular opening of $\Phi = \frac{\pi}{4(\gamma + 1)}$. To achieve super-renormalizability, the degrees of the polynomials in the definitions of $ H_0(\Box)$ and $H_2(\Box)$ must be equal. In the following discussion, we will not focus on the specific forms of the form factors $F_{i}(\Box)$ and will disregard the contributions from the interaction term $V$. 

\section{Ghost-free condition in de-Sitter spacetime}
The ghost-free condition is fundamental to ensure the stability and physical viability of field theories and gravitational models. The ghosts (unphysical degrees of freedom with negative kinetic energy) can lead to instability and unbounded negative energy states, making a theory unphysical. Therefore, satisfying this condition ensures that all propagating degrees of freedom possess positive kinetic terms, which is essential for maintaining vacuum stability and consistent predictions.
Moreover, while GR was initially formulated without ghost modes, modifications and extensions of GR that are designed to tackle various cosmological and astrophysical challenges often introduce higher derivative terms or additional fields that can inevitably lead to ghost instabilities. Consequently, constructing viable alternative theories of gravity, such as $f(R)$ gravity, higher derivative gravity, or nonlocal gravity theories, demands a meticulous formulation to prevent the emergence of these instabilities \cite{delhom2022instabilitiesfieldtheorieslecture,Arbuzova:2019xti,2014PhLB..739..279A,Felice_2006,DeFelice:2006pg}.

Additionally, the de-Sitter solution represents an exponentially expanding universe, which is also essential for the early inflationary phase and the current accelerated expansion \cite{PhysRevD.109.023515,Cicoli:2023opf,Vazquez:2018qdg}. There are numerous compelling theoretical and observational reasons to investigate gravitational theories around de-Sitter rather than the flat spacetime. Primarily, the true gravitational vacuum in quantum field theory remains ill-defined, as illustrated by the cosmological constant problem \cite{RevModPhys.61.1,Sol_Peracaula_2022,PhysRevD.106.083510}. Minkowski spacetime may not represent the true gravitational vacuum. In some theories, this state might even decay, potentially through the spontaneous production of ghost particles, as observed in higher-derivative gravity models. Consequently, perturbative calculations around such false vacuum can exhibit rapid divergence and lack reliability due to the presence of various types of instabilities. A possible approach is to identify a different vacuum state and examine quantum perturbations around this new vacuum. The de-Sitter spacetime is notable in that they maintain a number of local generators consistent with the Poincare group, similar to flat spacetime. On the other hand, incorporating background spacetimes of constant curvature represents a relatively minor modification that can be handled precisely without substantial computational effort. Thus, it provides an intriguing opportunity to explore the perturbative implications of the theory within the context of de-Sitter background. In this section, we will examine the ghost-free condition in the context of de-Sitter solution.
\subsection{Equations of Motion and on-shell condition}
The equations of motion for action \eqref{1} can be obtained by directly varying the action, resulting in the following expression \cite{Koshelev:2013lfm,Dimitrijevic:2015eaa}
\be
\label{6}
\begin{split}
&\(M^2_p+2\lambda F_{0}(\Box)R\)G^{\mu}_{\nu}=-\Lambda\delta^{\m}_{\n}-\frac{\lambda}{2}RF_{0}(\Box)R\delta^{\mu}_{\nu}+2\lambda\(\na^{\mu}\na_{\n}-\delta^{\m}_{\n}\Box\)F_{0}(\Box)R\\
&-2\lambda R^{\m}_{\beta}F_2(\Box)R^{\beta}_{\n}+\frac{\lambda}{2}\delta^{\m}_{\n}R^{\al}_{\beta}F_2(\Box)R^{\beta}_{\al}+\\
&+2\lambda\(\na_{\rho}\na_{\n}F_2(\Box)R^{\m\rho}-\frac{1}{2}\Box F_2(\Box)R^{\m}_{\n}-\frac{1}{2}\delta^{\m}_{\n}\na_{\sigma}\na_{\rho}F_2(\Box)R^{\sigma\rho}\)\\
&+\lambda K^{\m}_{1\n}-\frac{\lambda}{2}\delta^{\m}_{\n}\(K^{\sigma}_{1\sigma}+\bar{K}_1\)+\lambda K^{\m}_{2\n}-\frac{\lambda}{2}\delta^{\m}_{\n}\(K^{\sigma}_{2\sigma}+\bar{K}_2\)+2\lambda \Delta^{\m}_{\n},
\end{split}
\ee
with
\be
\1\{\begin{split}
\label{7}
&K^{\m}_{1\n}=\sum_{n=1}^{\infty} f_{0 n} \sum_{l=0}^{n-1} \partial^\mu R^{(l)} \partial_\nu R^{(n-l-1)}, \bar{K}_1=\sum_{n=1}^{\infty} f_{0 n} \sum_{l=0}^{n-1} R^{(l)} R^{(n-l)} ,\\
&K^{\m}_{2\n}=\sum_{n=1}^{\infty} f_{2_n} \sum_{l=0}^{n-1} \nabla^\mu R^{(l)}{ }_\beta^\alpha \nabla_\nu R^{(n-l-1)^\beta}{ }_\alpha, \bar{K}_2=\sum_{n=1}^{\infty} f_{2_n} \sum_{l=0}^{n-1} R^{(l)}{ }_\beta^\alpha R^{(n-l)}{ }_\alpha^\beta ,\\
&\Delta_\nu^\mu=\sum_{n=1}^{\infty} f_{2 n} \sum_{l=0}^{n-1} \nabla_\beta\left[R^{(l)}{ }_\gamma^\beta \nabla^\mu R^{(n-l-1)}{ }_\nu^\gamma-\nabla^\mu R^{(l)}{ }_\gamma^\beta R^{(n-l-1)}{ }_\nu^\gamma\right].
\end{split}\2.
\ee
where $G_{\m\n}$ is the Einstein tensor, $R^{(n)} \equiv \Box^nR$, and $R^{(n)}_{\al\beta} \equiv \Box^nR_{\al\beta}$. 
Upon analyzing Eq.\eqref{6}, we observe that $ F_0(\Box)= F_2(\Box) = 0$ corresponds to the canonical equations of motion for GR with a cosmological constant. When $F_0(\Box) = 1$, $F_2(\Box)=0$, and $\Lambda=0$, this extends to the specific case of local $f(R)$ gravity, commonly referred to as $R^2$ gravity. In addition, form factors $F_i(\Box)$ $(i=0,2)$ uniquely characterizes higher-derivative (potentially non-local) modifications of gravity. For the convenience of our discussion, we contract the indices of Eq.\eqref{6} and obtain the trace equation
\be
\label{8}
\begin{split}
& -M_P^2 R=-4 \Lambda-6 \lambda \square \mathcal{F}_0(\square) R-\lambda\left(K_1+2 \bar{K}_1\right)\\
& -\lambda \square F_2(\square) R-2 \lambda \nabla_\rho \nabla_\mu F_2(\square) R^{\mu \rho}-\lambda\left(K_2+2 \bar{K}_2\right)+2 \lambda \Delta.
\end{split}
\ee
The scalar parts denote the tensors undergoing self-contraction, and it is noteworthy that the terms involving form factors on both sides of the equation can be interchanged due to the property of integration by parts.

Furthermore, Ref.\cite{Calcagni_2018,Briscese:2018bny,Briscese:2019rii} have established that solutions of GR are also solutions within nonlocal gravity models, suggesting that de-Sitter solution should similarly be valid in this framework. By substituting $R = \text{constant}$ into Eq.\eqref{8}, we immediately derive the trace equation $M_p^2 R = 4 \Lambda$ (on-shell condition), which is consistent with GR. This also clearly demonstrates that the de-Sitter solution is indeed a valid solution within the nonlocal model. Subsequently, Ref.\cite{Calcagni_2017,Calcagni_2018,Calcagni_2008,Briscese:2018bny,Briscese:2019rii} have also elaborated on the stability of de-Sitter solution from both linear and nonlinear perturbative perspectives, though these results are perturbative in nature. Although their proofs are quite clear, the introduction of perturbations still leads to considerable computational complexity. In the following section, we will employ a novel proof strategy to address the stability of the de-Sitter solution.
\subsection{No-ghost excitations}
The de-Sitter space is maximally symmetric, which greatly facilitates technical calculations. Analyzing excitations in this context is instrumental for calculating the power spectrum of cosmological perturbations \cite{Craps:2014wga,Koshelev:2016xqb,Biswas:2012bp}. Utilizing the properties of the maximal symmetry group, we employ the covariant mode decomposition as introduced in \cite{D_Hoker_1999, Fradkin:1983mq}.
\be
\label{9}
h_{\m\n}=h^{\bot}_{\m\n}+\bar{\na}_{\m}A^{\bot}_{\n}+\bar{\na}_{\n}A^{\bot}_{\m}+\(\na_{\m}\na_{\n}-\frac{1}{4}\bar{g}_{\m\n}\Box\)B+\frac{1}{4}\bar{g}_{\m\n}h.
\ee
Here, $\bar{g}_{\m\n}$ and $\bar{\na}_{\m}$ denote the background metric and operator, respectively. The tensor $h^{\bot}_{\mu\nu}$ is transverse and traceless (spin-2), satisfying $\bar{\nabla}^{\mu} h^{\bot}_{\mu\nu} = \bar{g}^{\mu\nu} h^{\bot}_{\mu\nu} = 0$. The vector $A^{\bot}_{\mu}$ is transverse (spin-1) with $\bar{\nabla}^{\mu} A^{\bot}_{\mu} = 0$. Both $B$ and $h$ are scalars (spin-0), with the operator acting on $B$ being traceless. From the viewpoint of group representation theory, modes of different spins don't mix at the linearized level, allowing for their independent analysis.

Specifically, these fields encompass six physical states. The three gauge degrees of freedom reduce the spin-2 field to the two helicity states of a graviton, while an additional gauge freedom reduces the vector field to its two transverse spin-1 helicity states. Furthermore, it has been shown in Ref.\cite{Biswas:2016egy} that the contributions from $A^{\bot}_{\m}$ and $\na_{\m}\na_{\n}B$ are absent in action \eqref{1}. As a result, we can express the decomposition as
\be
\label{10}
h_{\m\n}=h^{\bot}_{\m\n}-\frac{1}{4}\bar{g}_{\m\n}\phi \quad\(\phi\equiv \bar{\Box} B-h\).
\ee

After extensive calculations, the final variational result can be expressed as
\be
\1\{\begin{split}
\label{11}
&\delta^2S(h^{\bot}_{\m\n})=\frac{1}{4}\int\dif^4x\sqrt{-g}h^{\bot}_{\m\n}\(\bar{\Box}-\frac{\bar{R}}{6}\)\(\frac{M^2_p}{2}+\lambda
f_{00}\bar{R}+\frac{\lambda}{4}f_{20}\bar{R}+\frac{\lambda}{2}F_2(\bar{\Box})\(\bar{\Box}-\frac{\bar{R}}{6}\)\)h^{\bot\m\n} ,\\
&\delta^2S(\phi)=-\frac{3}{32}\int\dif^4x\sqrt{-g}\phi\(\bar{\Box}+\frac{\bar{R}}{3}\)\( \frac{M^2_p}{2}+\lambda f_{00}\bar{R}+\frac{\lambda}{4}f_{20}\bar{R}-\lambda F_0(\bar{\Box})\(3\bar{\Box}+\bar{R}\)\right.\\
&\left.-\frac{\lambda}{4}F_2\(\bar{\Box}+\frac{2\bar{R}}{3}\)\bar{\Box}-\frac{\lambda}{4}F_{2}(\bar{\Box})\(3\bar{\Box}+\bar{R}\)  \)\phi,
\end{split}\2.
\ee
where the on-shell condition has been applied, and detailed calculations are provided in Appendix A. Moreover, we see that the nonlocal model exhibits a structure similar to GR, with the primary distinction being the inclusion of nonlocal operators. Notably, under certain parameter choices, the inverse of the quadratic Lagrangians exhibits the same propagator in Minkowski spacetime \cite{Biswas_2012,Krasnikov:1987yj,Tomboulis:1997gg,Koshelev:2017ebj,Modesto:2011kw}.

The condition for ensuring the absence of ghosts is (1): there should be no additional zeros in the spin-2 quadratic form beyond those present in pure GR. (2): the spin-0 quadratic form may include at most one additional zero, denoted as $\Box = m^2$ with positive $m^2$ to ensure it is not a tachyon. The presence of an extra pole (scalaron), typically associated with the Brans-Dicke scalar mode commonly found in pure $f(R)$ gravity, can be utilized to drive the inflationary period in the early universe. To satisfy these two conditions, we need to impose constraints on Eq.\eqref{11}, which must fulfill
\be
\1\{\begin{split}
\label{12}
&T_1(\bar{\Box})\equiv 1+\frac{2\lambda}{M^2_p} f_{00}\bar{R}+\frac{\lambda}{2M^2_p} f_{20}\bar{R}+\frac{\lambda}{M^2_p}F_2(\bar{\Box})\(\bar{\Box}-\frac{\bar{R}}{6}\)=e^{-2h_1(\bar{\Box})},\\
&T_2(\bar{\Box})\equiv1+\frac{2\lambda}{M^2_p} f_{00}\bar{R}+\frac{\lambda}{2M^2_p} f_{20}\bar{R}-\frac{2\lambda}{M^2_p} F_0(\bar{\Box})\(3\bar{\Box}+\bar{R}\)\\
&-\frac{\lambda}{2M^2_p}F_2\(\bar{\Box}+\frac{2\bar{R}}{3}\)\bar{\Box}-\frac{\lambda}{2M^2_p}F_{2}(\bar{\Box})\(3\bar{\Box}+\bar{R}\)=\(1-\frac{\bar{\Box}}{m^2}\)^{g}e^{-2h_2(\bar{\Box})}.
\end{split}\2.
\ee
Where $h_1(\bar{\Box})$ and $h_2(\bar{\Box})$ being entire functions, resulting in no roots from the exponential factor. The factor $g$ can only take the values 0 or 1. It is imperative to recognize that the restriction on the form factor derived above should be considered a universal requirement, applicable to all stages of the background's evolution, to ensure the theory remains consistently well-behaved. Redefining the field $\tilde{h}^{\bot}_{\m\n}\equiv\frac{M_p}{2}e^{-h_1(\bar{\Box})}h^{\bot}_{\m\n}$ and $\tilde{\phi}\equiv M_p\sqrt{\frac{3}{32}}e^{-h_2(\bar{\Box})}\phi$, we finally get
\be
\1\{\begin{split}
\label{13}
&\delta^2S(\tilde{h}^{\bot}_{\m\n})=\frac{1}{2}\int\dif^4x\sqrt{-g}\tilde{h}^{\bot}_{\m\n}\( \bar{\Box}-\frac{\bar{R}}{6} \)\tilde{h}^{\bot\m\n},\\
&\delta^2S(\tilde{\phi})=-\frac{1}{2}\int\dif^4x\sqrt{-g}\tilde{\phi}\(\bar{\Box}+\frac{\bar{R}}{3}\)\(1-\frac{\bar{\Box}}{m^2}\)^{g}\tilde{\phi}.
\end{split}\2.
\ee
It is evident that $g = 0$ corresponds to GR, whereas $g=1$ signifies the presence of a scalar particle that is not a ghost (The coefficient of the kinetic term is -1). From the perspective of nonlocal gravitational spectrum analysis, this model is nearly equivalent to GR and may introduce only one additional physical degree of freedom with $m^2 > 0$, indicating the stability of the de-Sitter solution. Additionally, we can also ascertain that this model satisfies unitarity by analyzing spectrum within de-Sitter background.

Subsequently, since $h_1(\bar{\Box})$ and $h_2(\bar{\Box})$ are two unknown functions, it is necessary to explicitly derive the analytic expressions of the form factors. Following the approach of Ref.\cite{Koshelev:2017ebj}, we impose the conditions $F_0(0)+\frac{1}{4}F_2(0)=0$, which are equivalent to setting $f_{00}+\frac{1}{4}f_{20}=0$. It is worth noting that this choice is consistent with the requirements of $H(z)$, although it is not unique and is adopted merely for convenience in solving. Ultimately, we obtain the general expressions for the form factors as

\be
\1\{\begin{split}
\label{14}
& F_0(\bar{\Box})=-\frac{M^2_p}{6\lambda}\( \frac{\(1-\frac{\bar{\Box}}{m^2}\)^g e^{H_0(\Box)}-1 }{\bar{\Box}+\frac{\bar{R}}{3}} \)-\frac{M^2_p}{4\lambda}\( \frac{e^{H_2\((\bar{\Box}-\frac{\bar{R}}{6})(\bar{\Box}-\frac{\bar{R}}{3})\)}-1}{\bar{\Box}-\frac{\bar{R}}{6}}  \) \\
&-\frac{M^2_p}{12\lambda}\bar{\Box}\( \frac{e^{H_2\((\bar{\Box}+\frac{\bar{R}}{2})(\bar{\Box}+\frac{\bar{R}}{3})\)}-1   }{\(\bar{\Box}+\frac{\bar{R}}{3}\)\(\bar{\Box}+\frac{\bar{R}}{2}\)}  \),     \\
& F_2(\bar{\Box})= \frac{M^2_p}{\lambda} \(\frac{  e^{H_2\( (\bar{\Box}-\frac{\bar{R}}{6})(\bar{\Box}-\frac{\bar{R}}{3}) \)}-1 }{\bar{\Box}-\frac{\bar{R}}{6}}   \).
\end{split}\2.
\ee
Here, $H_i(\bar{\Box})$ with $i=0,2$ are defined in Eq.\eqref{5}. Specifically, $H_0(\bar{\Box})=\frac{1}{2}\( \Gamma(0,p_0(\bar{\Box})^2)\right.\\
\left.+\gamma_E+\log{p_0(\bar{\Box})^2}\)$, where $p_0(\bar{\Box})$ can be chosen as  polynomial of $\bar{\Box}\(\bar{\Box}+\frac{\bar{R}}{3}\)$ to guarantee the absence of poles in the function $F_0(\bar{\Box})$. The function $H_2\((\bar{\Box}+\frac{\bar{R}}{2})(\bar{\Box}+\frac{\bar{R}}{3})\)$ is defined as $H_2\(\(\bar{\Box}-\frac{\bar{R}}{6}\)\(\bar{\Box}-\frac{\bar{R}}{3}\)\)|_{\bar{\Box}\to\bar{\Box}+\frac{2\bar{R}}{3}}$, accompanied by the polynomial $p_2\( \(\bar{\Box}-\frac{\bar{R}}{6}\)\(\bar{\Box}-\frac{\bar{R}}{3}\)\)$. We have already provided the most general solution for the form factors. Compared to the condition Eq.(12), it follows that the two unknown functions can be expressed as
\be
\1\{\begin{split}
\label{15}
&T_1(\bar{\Box})\equiv e^{-2h_1(\bar{\Box})}=e^{H_2\(\(\bar{\Box}-\frac{\bar{R}}{6}\)\(\bar{\Box}-\frac{\bar{R}}{3}\)\)  },\\
&T_2(\bar{\Box})\equiv\(1-\frac{\bar{\Box}}{m^2}\)^{g}e^{-2h_2(\bar{\Box})}=\(1-\frac{\bar{\Box}}{m^2}\)^{g}e^{H_{0}(\bar{\Box})}.
\end{split}\2.
\ee
Specifically, we consider two special cases. When $F_0(\bar{\Box}) \neq 0$ and $F_2(\bar{\Box}) = 0$, the two conditions reduce to
\be
\1\{\begin{split}
\label{16}
&T_1(\bar{\Box})= 1>0,\\
&T_2(\bar{\Box})=1-\frac{2\lambda}{M^2_p} F_0(\bar{\Box})\(3\bar{\Box}+\bar{R}\)=\(1-\frac{\bar{\Box}}{m^2}\)^{g}e^{-2h_2(\bar{\Box})}\\
&\Rightarrow F_0(\bar{\Box})=\frac{1-\(1-\frac{\bar{\Box}}{m^2}\)^{g}e^{-2h_2(\bar{\Box})}}{\frac{2\lambda\bar{R}}{M^2_p}\(1+\frac{3\bar{\Box}}{\bar{R}}\)}.
\end{split}\2.
\ee
The above conditions are a special case of Eq.\eqref{14}. Additionally, for $F_0(\bar{\Box})=0$ and $F_2(\bar{\Box})\neq0$, $F_2(\bar{\Box})$ provides only trivial solution ($F_2(\bar{\Box})=0$), which are not relevant to our analysis. From Eq.\eqref{15}, it follows that to achieve consistency on both sides, $h_2(0)$ must be zero, which aligns with the definition of a weak nonlocal model \cite{Koshelev:2017ebj}.

However, two issues remain to be addressed. The presence of a pole in the denominator contradicts the assumption that the form factors are entire functions; This issue is naturally resolved by the result of Eq.\eqref{14}, which demonstrates that the form factors are free of poles. Additionally, the results indicate that linking a cosmological constant $\Lambda$ to the form factors present issues. In other words, $\Lambda$ is associated with $F_0(\bar{\Box})$ and $F_2(\bar{\Box})$. To avoid the emergence of different form factors corresponding to different values of $\Lambda$, we define the form factors defined in Eq.\eqref{14} as the initial model choice (with $\bar{R}$ replaced by $\frac{4\Lambda}{M^2_p}$), in a manner analogous to Eq.\eqref{2}. This prescription effectively resolves the issue.

\section{Stability analysis}
The stability is also crucial in studying the evolution of fluctuation within cosmological models. If de-Sitter space is stable, it indicates that the universe can maintain this state despite small perturbations, thereby providing a robust foundation for inflationary models. Conversely, an unstable de-Sitter solution may signal a transition to a different cosmological phase, significantly impacting the universe's fate. Additionally, the stability can understand quantum gravity, as it tests the consistency of these theoretical frameworks with observed cosmological phenomena. Thus, investigating the stability of de-Sitter solutions deepens our understanding of cosmic dynamics.

However, studies on the stability of the de-Sitter solution in the context of nonlocal gravity remain limited. Existing works have primarily shown that the solutions of GR also hold for nonlocal gravity, with both models exhibiting identical stability properties. In other words, the stability of the de-Sitter solution in GR directly governs the stability of the corresponding solution in nonlocal gravity \cite{Briscese:2019rii}. Nevertheless, the stability of the de-Sitter solution in GR has long been an unresolved issue. It has been suggested for some time that the de-Sitter geometry may be unstable under quantum fluctuations \cite{Polyakov:1982ug,Antoniadis:1991fa,Tsamis:1992sx,Ford:1984hs,Mottola:1984ar,Antoniadis:1985pj,Mazur:1986et,Matsui:2019tlf,Polyakov:2007mm,Krotov:2010ma}. Additionally, some perspectives argue that the de-Sitter solution is stable \cite{Abbott:1981ff,Bruni:2001pc,Moreau:2018lmz}. In particular, using the Wilsonian renormalization group, it has been suggested that unbounded loop corrections in the deep infrared are ultimately screened by nonperturbative effects, which in turn stabilize the geometry \cite{Moreau:2018lmz}. These studies focus on the analysis of quantum fields within the de-Sitter background. The fact that the stability in GR aligns with that in nonlocal gravity makes the stability analysis in nonlocal gravity more challenging. To tackle this problem, we introduce a novel approach. Rather than relying on the stability analysis of GR, we establish the equivalence between nonlocal gravity and higher-derivative gravity through recursion relations. This equivalence allows us to demonstrate the stability of nonlocal gravity in the de-Sitter solution by employing higher-derivative gravity. It should be emphasized that our analysis is carried out strictly at the classical level, without incorporating quantum effects, and is restricted to the vacuum case, excluding any matter fields.


\subsection{Recursion relation for equation of motion}
Due to the complexity of expression \eqref{6}, constructing a general solution remains a challenging endeavor. However, it is noteworthy that significant progress has been achieved in the nonlocal case by adopting a simplifying ansatz
\be
\label{17}
\Box R=r_1R+r_2\quad (r_1\neq0)
\ee
in the absence of $F_2(\Box)$ and $F_4(\Box)$ \cite{Biswas:2005qr}. Here, $r_1$ and $r_2$ are constants. Indeed, it turns out that
\be
\1\{\begin{split}
\label{18}
&\square^n R=r_1^n\left(R+r_2 / r_1\right) \quad\text { for } n>0, \\
& F_0(\square) R=F_0(r_1)+\frac{r_2}{r_1}\left(F_0(r_1)-f_{00}\right) .
\end{split}\2.
\ee

By substituting these relations into the equations of motion (EOM) \eqref{6} and performing additional algebraic manipulations, we obtain the same result as in Ref.\cite{Biswas:2005qr}. Specifically, a solution of Eq.\eqref{17} is also a solution to the full nonlocal EOM \eqref{8}, provided that specific algebraic conditions
\be
\label{19}
F'\left(r_1\right)=0,   \quad\frac{r_2}{r_1}\left(F_0(r_1)-f_{00}\right)=-\frac{M_P^2}{2 \lambda}+3 r_1 F_0(r_1), \quad 4 r_1 \Lambda=-r_2 M_P^2 .
\ee
Here $F'\left(r_1\right)$ denotes the first derivative with respect to its parameter. The form of cosmological solutions in the nonlocal model has been established \cite{Biswas:2012bp}. There are three distinct cosmological scenarios:

$\bullet$ Cyclic universe scenario: For $\Lambda < 0$, $r_1 > 0\Longrightarrow r_2 > 0$, the universe undergoes successive bounces and turnarounds.

$\bullet$ Bouncing universe scenario: For $\Lambda > 0$, $r_1 < 0\Longrightarrow r_2 > 0$, the universe transitions from a phase of contraction to a phase of super-inflating expansion.

$\bullet$ Geodesically complete bouncing universe: For $\Lambda >0$, $r_1 > 0\Longrightarrow r_2 < 0$, this scenario represents a non-singular bounce that completes an inflationary phase and admits constant curvature vacuum solutions, such as the de-Sitter or Minkowski solutions, depending on the value of the cosmological constant.


Furthermore, Ref.\cite{Koshelev:2013lfm} presents a recursion relation for handling $F_2(\Box)$, which reads
\be
\1\{\begin{split}
\label{20}
&\Box^{n} \tilde{G}^{\m}_{\n}=s^{n}_{1} S^{\m}_{1\n}-s^{n}_{2}  S^{\m}_{2\n} \quad\text{for }  n \geqslant 0,\\
&F_2(\square) R_\nu^\mu=F_2\left(s_1\right) S_{1\n}^\mu-F_2\left(s_2\right) S_{2\n}^\mu+\frac{1}{4} \delta_\nu^\mu F_2(\square) R,
\end{split}\2.
\ee
where $S^{\m}_{1\n}\equiv\frac{\square \tilde{G}_\nu^\mu-s_2 \tilde{G}_\nu^\mu}{6 \sigma^2}$, $S^{\m}_{2\n}\equiv\frac{ \square \tilde{G}_\nu^\mu-s_1 \tilde{G}_\nu^\mu}{6 \sigma^2}$, $\tilde{G}^{\m}_{\n}\equiv R^{\m}_{\n}-\frac{1}{4}\delta^{\m}_{\n}R$, and $s_1$, $s_2$, and $\sigma$ are parameters. These relations yield
\be
\1\{\begin{split}
\label{21}
&K_{1\n}^\mu=F'_0\left(r_1\right) \partial^\mu R \partial_\nu R,\\
&\bar{K}_1=r_1 F'_0\left(r_1\right) R^2+2 r_2 F'_0\left(r_1\right) R-\left(F_0\left(r_1\right)-f_{00}\right) R \frac{r_2}{r_1}
+\frac{r_2^2}{r_1} F'_0\left(r_1\right)-\left(F_0\left(r_1\right)-f_{00}\right)\left(r_2 / r_1\right)^2,\\
&K_{2\nu}^\mu=F'_2\left(s_1\right) \nabla^\mu S_1{ }_\beta^\alpha \nabla_\nu S_1{ }_\alpha^\beta+F'_2\left(s_2\right) \nabla^\mu S_2{ }_\beta^\alpha \nabla_\nu S_2{ }_\alpha^\beta+\frac{1}{4}F'_2\left(r_1\right) \partial^\mu R \partial_\nu R \\
&-\frac{F_2\left(s_1\right)-F_2\left(s_2\right)}{s_1-s_2}\left(\nabla^\mu S_1{ }_\beta^\alpha \nabla_\nu S_2{ }_\alpha^\beta+\nabla^\mu S_2{ }_\beta^\alpha \nabla_\nu S_1{ }_\alpha^\beta\right) ,\\
&\bar{K}_2=s_1 F'_2\left(s_1\right) S_{1\beta}^{\alpha} S_{1\al}^{\beta}+s_2 F'_2\left(s_2\right) S_{2\beta}^\alpha S_{2\alpha}^\beta-\frac{F_2\left(s_1\right)-F_2\left(s_2\right)}{s_1-s_2}\left(s_1+s_2\right) S_1{ }_\beta^\alpha S_{2\al}^\beta \\
&+\frac{1}{4}\left(r_1 F'_2\left(r_1\right) R^2+2 r_2 F'_2\left(r_1\right) R-F_2\left(r_1\right) R \frac{r_2}{r_1}+\frac{r_2^2}{r_1}
F'_2\left(r_1\right)-F_2\left(r_1\right)\left(r_2 / r_1\right)^2\right) ,\\
&\Delta_\nu^\mu= \nabla_\beta \left[ F'_2\left(s_1\right)\left(S_{1\epsilon}^\beta \nabla^\mu  S_{1\nu}^\epsilon-\nabla^\mu S_{1\epsilon}^\beta S_{1\n}^\epsilon\right)
+F'_2\left(s_2\right)\left(S_{2\epsilon}^\beta \nabla^\mu S_{2\n}^\epsilon-\nabla^\mu S_{2\epsilon}^\beta S_{2\n}^\epsilon\right)\right. \\
& \left.-\frac{F_2\left(s_1\right)-F_2\left(s_2\right)}{s_1-s_2}\left(S_{1\epsilon}^\beta \nabla^\mu S_{2\nu}^\epsilon-\nabla^\mu S_{1\epsilon}^{ \beta} S_{2\n}^\epsilon+S_{2\epsilon}^\beta \nabla^\mu S_{1\nu}^\epsilon-\nabla^\mu S_{2\epsilon}^{ \beta} S_{1\n}^\epsilon\right)\right].
\end{split}\2.
\ee

To cancel the terms involving $\Box\tilde{G}^{\m}_{\n}$, $\(\Box\tilde{G}^{\m}_{\n}\)^2$, $R^2$ and $\(\partial_\m R\)^2$ in the trace equation, the following conditions must be imposed
\be
\1\{\begin{split}
\label{22}
&F'_2\left(s_1\right)=F'_2\left(s_2\right)=F_2\left(s_1\right)-F_2\left(s_2\right)=0,\\
&F'_0\left(r_1\right)+\frac{1}{4} F'_2\left(r_1\right)=0 .
\end{split}\2.
\ee
As outlined in the equation above, we will proceed with the trace equation \eqref{8} . In the case of traceless matter (or in the absence of matter), it simplifies to
\be
\label{23}
\begin{aligned}
& -M_P^2 R=-4 \Lambda-6 \lambda\left( F_0\left(r_1\right)+\frac{1}{4} F_2\left(r_1\right)   \right)\left(r_1 R+r_2\right)-\frac{\lambda}{2} \mathcal{F}_2\left(s_1\right)\left(r_1 R+r_2\right)+ \\
& +2 \lambda\left(F_0\left(r_1\right)+\frac{1}{4}F_2\left(r_1\right)-f_{00}-\frac{1}{4}f_{20}\right) \frac{r_2}{r_1}\left(R+\frac{r_2} { r_1}\right).
\end{aligned}
\ee
Thus we solve it by imposing
\be
\1\{\begin{split}
\label{24}
&\frac{M_P^2}{r_1}-\frac{\lambda}{2} F_2\left(s_1\right)-6 \lambda\left(F_0\left(r_1\right)+\frac{1}{4} F_2\left(r_1\right)\right)+2 \lambda\left(F_0\left(r_1\right)+\frac{1}{4}F_2\left(r_1\right)-f_{00}-\frac{1}{4}f_{20}\right) \frac{r_2}{r^2_1}  =0 ,\\
&-\frac{M_P^2}{4} \frac{r_2}{r_1}  =\Lambda.
\end{split}\2.
\ee

We obtain that this result is simply an extension of Eq.\eqref{19}. In other words, solving the complex field equation is equivalent to solving Eq.\eqref{17} and \eqref{20} while simultaneously accounting for the constraint conditions specified by Eq.\eqref{22} and \eqref{24}. Similarly, Ref.\cite{Koshelev:2013lfm} provides the conditions for the existence of the de-Sitter solution, which are consistent with those of the geodesically complete bouncing universe discussed above. We will also use this condition in the following subsection to prove the stability of the de-Sitter solution.

Furthermore, it is noteworthy that the trace equation does not account for the entire system of gravitational field equations. For an Friedmann-Robertson-Walker (FRW) background, it is essential to also consider the (00)-component of the Einstein equations in the presence of matter. To ensure that the energy density $\rho$ remains positive, an additional constraint $F_0\left(r_1\right)+\frac{1}{4} F_2\left(r_1\right)+\frac{1}{12}F_2\left(s_2\right)<0$ is necessary \cite{Koshelev:2013lfm}. This constraint ensures that the matter contribution maintains positive energy and avoids the presence of ghosts.

\subsection{Equivalence with higher-order derivative model }
On the other hand, we consider a general higher-order derivative theory, with its action given by
\be
\label{25}
S=\int \dif^4x\sqrt{-g}\[\frac{\tilde{M}^2_P}{2}R+\frac{\lambda}{2}\(\tilde{f}_{00}R^2+\tilde{f}_{20}R_{\m\n}R^{\mu\nu}\)-\tilde{\Lambda}  \],
\ee
where $\tilde{M}_P$ and $\tilde{\Lambda}$  do not correspond to the actual Planck constant or cosmological constant. The parameters are denoted here with hats, and Eq.\eqref{25} represents an effective form of Eq.\eqref{1} in which the non-local operators are reduced to constant terms. However, to demonstrate the equivalence of the two actions, we derive the EOM from the expression above and take the trace, resulting in
\be
\label{26}
\tilde{M}^2_PR=4\tilde{\Lambda}+6\lambda\tilde{f}_{00}\Box R+2\lambda\tilde{f}_{20}\Box R.
\ee
Subsequently, by applying condition \eqref{17} and comparing it with Eq.\eqref{23}, we can derive the constraint conditions for the equivalence of the two models
\be
\1\{\begin{split}
\label{27}
&\tilde{M}^2_P=M^2_P+2 \lambda\left(F_0\left(r_1\right)+\frac{1}{4}F_2\left(r_1\right)-f_{00}-\frac{1}{4}f_{20}\right) \frac{r_2}{r_1},\\
&\tilde{\Lambda}=\Lambda-\frac{ \lambda}{2}\left(F_0\left(r_1\right)+\frac{1}{4}F_2\left(r_1\right)-f_{00}-\frac{1}{4}f_{20}\right) \frac{r^2_2}{r^2_1},\\
&\tilde{f}_{00}=F_0\left(r_1\right)+\frac{1}{4} F_2\left(r_1\right), \\
&\tilde{f}_{20}=\frac{1}{4} F_2\left(s_1\right).
\end{split}\2.
\ee

Therefore, we formulate the conditions under which the two actions become equivalent. Specifically, we demonstrate that the nonlocal model is equivalent to the higher-order derivative model, provided that the condition in Eq.\eqref{27} is satisfied, with the requirement being that the recursion relations in Eqs.\eqref{17} and \eqref{20} are concurrently satisfied.

Additionally, since the parameter $\tilde{M}_P$ might be negative, we seek to attribute physical significance to it by equating it with the Planck mass, which requires setting $F_0\left(r_1\right)+\frac{1}{4}F_2\left(r_1\right)-f_{00}-\frac{1}{4}f_{20}=0$ in Eq.\eqref{27}. This approach naturally results in $\Lambda=\tilde{\Lambda}$, implying that the de-Sitter solution of the two actions are consistent. Furthermore, since the de-Sitter solution has been shown to satisfy the constraint conditions and recursion relations in Ref.\cite{Koshelev:2013lfm,Biswas:2012bp,Biswas:2010zk}, we can naturally transition the investigation of its stability in the nonlocal model to an analysis within the framework of the higher-order derivative model.

In Appendix B, we demonstrate using the minimal superspace approach that for a general $f(R,R_{\m\n}R^{\m\n})$ model, the on-shell condition and the criterion for the stability of the de-Sitter solution are given by
\be
\1\{\begin{split}
\label{28}
&\bar{f}-\frac{\bar{R}}{2}\bar{f}_{R}-\frac{\bar{R}^2}{4}\bar{f}_{X} =0,\\
&\frac{\bar{f}_{R}+\frac{2}{3}\bar{R}\bar{f}_{X}}{3\bar{f}_{RR}+2\bar{f}_{X}+3\bar{R}\bar{f}_{RX}+\frac{3}{4}\bar{R}^2\bar{f}_{XX}}>\frac{\bar{R}}{3},
\end{split}\2.
\ee
where the subscripts on $f$ denote derivatives with respect to its arguments: $f_R=\frac{\partial f}{\partial R}$,  $f_X=\frac{\partial f}{\partial X}$, $f_{R R}=\frac{\partial^2 f}{\partial R^2}$,  $f_{XR}=\frac{\partial^2 f}{\partial R \partial X}$, and $f_{XX}=\frac{\partial^2 f}{\partial X^2}$ $\(X\equiv R_{\m\n}R^{\m\n}\)$. The bar on any quantity, as in the previous section, indicates that it is evaluated on the background. Subsequently, by substituting Eq.\eqref{25} into the on-shell condition, we can readily derive $M^2_P\bar{R}=4\Lambda$, which is equivalent to the trace equation \eqref{8} and \eqref{26} ($\bar{R}=\text{constant}$). Following this, substituting Eq.\eqref{25} into the stability condition yields
\be
\label{29}
\frac{\tilde{M}^2_P}{6\lambda\(\tilde{f}_{00}+\frac{1}{3}\tilde{f}_{20} \)}=\frac{M^2_P}{6\lambda\( F_0\left(r_1\right)+\frac{1}{4} F_2\left(r_1\right)+\frac{1}{12} F_2\left(s_1\right) \)}>0\Rightarrow r_1>0.
\ee

In the final step, we make use of Eq.\eqref{24} and take into account the imposed constraints. Based on our analysis, we conclude that the de-Sitter solution is stable in the absence of contributions from matter fields.

\section{conclusion and discussion}
In this paper, we first presented a general nonlocal gravity where the action includes quadratic forms of both the Ricci scalar and the Ricci tensor. It is worth noting that the action in Eq.\eqref{1} was not derived within the Weyl basis. This model introduced two specific form factors, $F_0(\Box)$ and $F_2(\Box)$, which are constructed to ensure the theory's super-renormalizability and unitarity. These form factors are entire functions and exhibit the same asymptotic behavior in the UV regime.

Subsequently, we analyzed the ghost-free condition of the nonlocal gravity within the de-Sitter background using perturbative approach. 
By absorbing the form factors into redefined fields, we discovered that the particle spectrum was nearly identical to GR, differing only in the presence of an additional scalar mode with a positive mass term $m$. We also provided general solutions for the form factors that ensure the ghost-free condition was maintained across arbitrary backgrounds. Furthermore, we examined the stability of the model under the de-Sitter solution from  perturbative perspective. This stability was vital because, by utilizing our previous method, we could demonstrate that the operator associated with the scalar mode has an eigenvalue greater than zero \cite{Feng_2024}.

We ultimately demonstrate that nonlocal gravity is fully equivalent to higher-derivative gravity when the constraint conditions and recurrence relations are satisfied, thereby establishing the equivalence expressed in Eq.\eqref{27}. Consequently, the stability analysis of the de-Sitter solution in the nonlocal model naturally transitions to the study of higher-order derivative gravity. Using the minimal superspace approach, we further derived the on-shell condition and stability constraints for a general $f(R,R_{\m\n}R^{\m\n})$ model in Appendix B. By substituting Eq.\eqref{25} into these constraints, we obtained the on-shell condition $M^2_PR=4\Lambda$ and the stability constraint $ r_1>0$ (condition for the validity of the de-Sitter solution). The result provided a proof of the stability of the de-Sitter solution within nonlocal gravity.

In the future, we will explore the dS/CFT correspondence. Although the AdS/CFT correspondence is well-established, a clear definition for the dS/CFT correspondence remains elusive, except when considering a nonlocal mapping between AdS and dS spaces, as discussed in \cite{Balasubramanian:2002zh}. We aim to apply the results from string theory and the AdS/CFT correspondence to nonlocal quantum gravity. We anticipate that exploring whether nonlocal gravity can provide new insights into the conceptual challenges associated with a potential dS/CFT correspondence will be particularly valuable.

\begin{acknowledgments}
This study was supported by the National Natural Science Foundation of China (Grant No. 12333008).
\end{acknowledgments}

\appendix

\section{The calculation of variation}
In this appendix, we focus on the second variation analysis of the action given by Eq.\eqref{1} in the de-Sitter solution. We know that the solution belongs to a maximally symmetric space which can be expressed as
\be
\1\{\begin{split}
\label{A1}
&R_{\m\n\sigma\rho}=\frac{R}{D(D-1)}\(g_{\m\sigma}g_{\n\rho}-g_{\m\rho}g_{\n\sigma}\),\\
&R_{\m\n}=\frac{R}{D}g_{\m\n},
\end{split}\2.
\ee
where $D$ denotes the spacetime dimension. In particular, the Weyl tensor $C_{\m\n\sigma\rho}=0$ in this background. Furthermore, we decompose the action into three parts. For the first part $S_0$, which can be defined as
\be
\label{A2}
S_0\equiv\int \dif^4x\sqrt{-g}\(\frac{M^2_P}{2}R-\Lambda\).
\ee
The variation result can be obtained by \cite{VanNieuwenhuizen:1973fi,N_ez_2005,Chiba_2005}
\be
\label{A3}
\delta^2 S_0=\int d^4 x \sqrt{|\bar{g}|} \frac{M_P^2}{2}\delta_0,
\ee
with
\be
\1\{\begin{split}
\label{A4}
&\delta_0\equiv\delta_{\mathrm{EH}}-\frac{2}{M_P^2} \Lambda \delta_g,\\
&\delta_{\mathrm{EH}}\equiv\left(\frac{1}{4} h_{\mu \nu} \bar{\square} h^{\mu \nu}-\frac{1}{4} h \bar{\square} h+\frac{1}{2} h \bar{\nabla}_\mu \bar{\nabla}_\rho h^{\mu \rho}+\frac{1}{2} \bar{\nabla}_\mu h^{\mu \rho} \bar{\nabla}_\nu h_\rho^\nu\right) \\
& +\left(h h^{\mu \nu}-2 h_\sigma^\mu h^{\sigma \nu}\right)\left(\frac{1}{8} \bar{g}_{\mu \nu} \bar{R}-\frac{1}{2} \bar{R}_{\mu \nu}\right)-\left(\frac{1}{2} \bar{R}_{\sigma \nu} h_\rho^\sigma h^{\nu \rho}+\frac{1}{2} \bar{R}_{\rho \nu \mu}^\sigma h_\sigma^\mu h^{\nu \rho}\right),\\
&\delta_g \equiv\frac{h^2}{8}-\frac{h_{\mu \nu}^2}{4} .
\end{split}\2.
\ee

Subsequently, substituting the tensor decomposition from Eq.\eqref{9} into the above expression yields
\be
\label{A5}
\delta_0(h^{\bot}_{\m\n},\phi)=\frac{1}{4}h^{\bot}_{\m\n}\(\bar{\Box}-\frac{\bar{R}}{6}\)h^{\bot\m\n}-\frac{1}{32}\phi\(3\bar{\Box}+\bar{R}\)\phi,
\ee
where the definition of $\phi$ is consistent with Eq.\eqref{10}. Furthermore, we analyze $S_1$ which can be represented as
\be
\label{A6}
S_1=\frac{\lambda}{2}\int \dif^4x\sqrt{-g}RF_0(\Box)R .
\ee
This variation has also been derived in Ref.\cite{Biswas:2016egy} and can be written as
\be
\label{A7}
\begin{split}
&\delta^2 S_1=\frac{\lambda}{2} \int d^4 x \sqrt{-\bar{g}}\left[2\left(\frac{h}{2} R^{(1)}+\frac{1}{2}\left(\frac{h^2}{8}-\frac{h_{\mu \nu} h^{\mu \nu}}{4}\right) \bar{R}+R^{(2)}\right) f_{00} \bar{R}+R^{(1)} F_0(\bar{\square}) R^{(1)}\right. \\
& \left.+\left(\frac{h}{2} \bar{R}+R^{(1)}\right) \delta\left(F_0(\bar{\square})\right) \bar{R}+\bar{R} \delta^2\left(F_0(\bar{\square})\right) \bar{R}+\frac{h}{2} \bar{R}\left(F_0(\bar{\square})-f_{00}\right) R^{(1)}+\bar{R} \delta\left(F_0(\bar{\square})\right) R^{(1)}\right],
\end{split}
\ee
with
\be
\1\{\begin{split}
\label{A8}
&R^{(1)}=\bar{\na}_{\m}\bar{\na}_{\n}h^{\m\n}-\bar{\Box} h-\bar{R}_{\m\n}h^{\m\n},\\
&R^{(2)}=\frac{1}{4}h_{\m\n}\bar{\Box} h^{\m\n}+\frac{1}{4}h\bar{\Box} h+\frac{1}{2}\bar{\na}_{\m}h^{\m\n}\bar{\na}^{\rho}h_{\n\rho}+\frac{1}{2}\bar{R}_{\m\n}h^{\m\alpha}h^{\n}_{\alpha}+\frac{1}{2}\bar{R}_{\m\n\sigma\rho}h^{\m\sigma}h^{\n\rho}.
\end{split}\2.
\ee

It can be demonstrated that the contribution from the second line of Eq.\eqref{A7} is zero. Incorporating the definition of $\delta_0$, the expression can be simplified to
\be
\label{A9}
\delta^2 S_{1}=\frac{\lambda}{2} \int d^4 x \sqrt{-\bar{g}}\left[2 f_{00} \bar{R} \delta_0+R^{(1)} F_0(\bar{\square})R^{(1)} \right] .
\ee
Using Eq.\eqref{10}, we can demonstrate that
\be
\label{A10}
\frac{\lambda}{2}R^{(1)} F_0(\bar{\square})R^{(1)}(h^{\bot}_{\m\n},\phi)=\frac{\lambda}{32}\phi   F_0(\bar{\square})   \(3\bar{\Box}+\bar{R}\)^2      \phi  .
\ee

It is clear that this term does not contribute to the transverse modes of the tensor $h_{\m\n}$. Ultimately, we turn our attention to the contribution from $S_2$, which follows
\be
\label{A11}
S_2=\frac{\lambda}{2}\int \dif^4x\sqrt{-g}R^{\m}_{\n}F_2(\Box)R^{\n}_{\m}.
\ee
where we select mixed upper and lower indices to facilitate subsequent manipulations. The variation of above equation is
\be
\label{A12}
\begin{split}
\delta^2 S_2&=\frac{\lambda}{2} \int d^4 x \sqrt{-\bar{g}}\[ 2f_{20}\bar{R}^{\m(2)}_{\n}\bar{R}^{\n}_{\m}+f_{20}h\bar{R}^{\m(1)}_{\n}\bar{R}^{\n}_{\m}+\bar{R}^{\m}_{\n}\bar{R}^{\n}_{\m}f_{20}\(\frac{1}{8}h^2-\frac{1}{4}h_{\m\n}h^{\m\n}\)+\bar{R}^{\m(1)}_{\n}F_2(\bar{\Box}) \bar{R}^{\n(1)}_{\m} \]\\
&=\frac{\lambda}{2} \int d^4 x \sqrt{-\bar{g}}\[ \frac{\delta_0}{2}f_{20}\bar{R}+\bar{R}^{\m(1)}_{\n}F_2(\bar{\Box}) \bar{R}^{\n(1)}_{\m}\].
\end{split}
\ee
Following the derivation scheme of $S_1$ and after extensive calculation, we can derive
\be
\1\{\begin{split}
\label{A13}
&\frac{\lambda}{2}R^{\m(1)}_{\n} F_2(\bar{\square})R^{\n(1)}_{\m}(h^{\bot}_{\m\n},\phi)=\frac{\lambda}{8}h^{\bot}_{\m\n}\(\bar{\Box}-\frac{\bar{R}}{6}\)F_2(\bar{\Box}) \(\bar{\Box}-\frac{\bar{R}}{6}\)h^{\bot\m\n}\\
&+\frac{\lambda}{128}\phi \(\( 3\bar{\Box}+\bar{R} \)  F_2(\bar{\square}) +\bar{\Box}F_2\(\bar{\square}+\frac{2\bar{R}}{3}\)   \)\(3\bar{\Box}+\bar{R}\)  \phi.
\end{split}\2.
\ee
\section{Stability of de-Sitter Solution in $f(R,R_{\m\n}R^{\m\n})$ gravity}
To determine whether this solution represents a stable minimum, we shift our focus to isotropic and homogeneous solutions using the spatially flat FRW metric
\be
\label{B1}
\dif s^2=-N(t)\dif t^2+a^2(t)\(\dif x^2+\dif y^2+\dif z^2\),
\ee
where $t$ denotes cosmic time, and $N(t)$ is an arbitrary lapse function that reflects the gauge freedom associated with the reparametrization invariance of the mini-superspace gravitational model. The curvature scalar and $R_{\m\n}R^{\m\n}$ can be given by
\be
\1\{\begin{split}
\label{B2}
&R=6\(\frac{\ddot{a}}{aN^2}+\frac{\dot{a}^2}{a^2N^2}-\frac{\dot{a}\dot{N}}{aN^3} \),\\
&X\equiv R_{\m\n}R^{\m\n}=\frac{12\dot{a}^4}{a^4N^4}-\frac{12\dot{a}^3\dot{N}}{a^3N^5}+\frac{12\dot{a}^2\dot{N}^2}{a^2N^6}+\frac{12\ddot{a}\dot{a}^2}{a^3N^4}-\frac{24\ddot{a}\dot{a}\dot{N}}{a^2N^5}+\frac{12\ddot{a}^2}{a^2N^4}.
\end{split}\2.
\ee
To work within the first-derivative gravitational framework, we introduce Lagrange multipliers $ y_1$ and $y_2$ to express the action as
\be
\label{B3}
\begin{split}
&S=\int \dif^3x\int \dif t Na^3\[f(R,X)-y_1\(R-6\(\frac{\ddot{a}}{aN^2}+\frac{\dot{a}^2}{a^2N^2}-\frac{\dot{a}\dot{N}}{aN^3} \)\) \right. \\
& \left.-y_2\(X-\(\frac{12\dot{a}^4}{a^4N^4}-\frac{12\dot{a}^3\dot{N}}{a^3N^5}+\frac{12\dot{a}^2\dot{N}^2}{a^2N^6}+\frac{12\ddot{a}\dot{a}^2}{a^3N^4}-\frac{24\ddot{a}\dot{a}\dot{N}}{a^2N^5}+\frac{12\ddot{a}^2}{a^2N^4} \)\)\].                             \end{split}
\ee
when we perform the variation with respect to $R$ and $X$, we obtain $y_1 = f_R $ and $y_2=f_X$. Furthermore, by substituting these results into the above equation and performing integration by parts, we derive the Lagrangian
\be
\label{B4}
\begin{split}
&L(a,\dot{a},R,\dot{R},N,\dot{N})=-\frac{6a\dot{a}^2f_R}{N}-\frac{6a^2\dot{a}\dot{R}f_{RR}}{N}+Na^3\(f-Rf_R-Xf_X\)\\
&+12f_X\( \frac{\dot{a}^4}{aN^3}-\frac{\dot{a}^3\dot{N}}{N^4}+  \frac{a\dot{a}^2\dot{N}^2}{N^5}+\frac{\dot{a}^2\ddot{a}}{N^3}-\frac{2a\dot{a}\ddot{a}\dot{N}}{N^4}+\frac{a\ddot{a}^2}{N^3}             \).
\end{split}
\ee
In this scenario, the Lagrangian incorporate $ a $, $R$, $N$, and their derivatives as independent variables. Three EOM can be derived, with two of them being independent. These equations form the basis of the analysis system. For this research, we set $N(t) = 1$. Therefore, the EOM corresponding to $R$ and $N$ are
\be
\1\{\begin{split}
\label{B5}
& \dot{H}=\frac{R}{6}-2H^2,\\
&\dot{R}=\frac{B(R,H)}{A(R,H)},
\end{split}\2.
\ee
with
\be
\1\{\begin{split}
\label{B6}
&B(R,H)\equiv f+\(6H^2-R\)f_R+24H^2\dot{H}\(12H^2-R\)f_{RX}-  \(12\dot{H}^2+24\dot{H}H^2+X\)f_X\\
&+48\dot{H}H^2f_{XX}\( 3H^2+2\dot{H} \)\( 12H^2-R \) ,                                 \\
&A(R,H)\equiv 4H\( 3H^2+2\dot{H}   \)\( -3f_{RX}+2\(3H^2-R\)f_{XX}\)  -6Hf_{RR}+4H\(3H^2-R\)f_{RX}\\
&-4Hf_X.
\end{split}\2.
\ee
Where  $H \equiv \frac{\dot{a}}{a}$ denotes the Hubble parameter. It is noteworthy that the Euler equation for $a$ does not need to be explicitly considered, as it can be derived from the two independent equations mentioned above.

The critical points $ \bar{R}$ and $ \bar{H}$, defined by $\dot{R} = 0$ and $\dot{H} = 0 $, are essential for examining the system's stability. Thus, the on-shell condition is equivalent to $\bar{R}=12\bar{H}^2$ and $\bar{f}-\frac{\bar{R}}{2}\bar{f}_{R}-\frac{\bar{R}^2}{4}\bar{f}_{X} =0$. Subsequently, the system is linearized at these critical points
\be
\label{B6}
\left(
\begin{array}{c}
\delta\dot{R}\\
\delta\dot{H}
\end{array}
\right)
=
\left(
\begin{array}{cc}
\bar{H}&\frac{-6\(\bar{f}_R+8\bar{H}^2\bar{f}_X \)}{3\bar{f}_{RR}+2\bar{f}_X  +36\bar{H}^2\(\bar{f}_{RX}+3\bar{H}^2\bar{f}_{XX} \)  }\\
\frac{1}{6}&-4\bar{H}
\end{array}
\right)
\left(
\begin{array}{c}
\delta R\\
\delta H
\end{array}
\right).
\ee
It is straightforward to demonstrate that these two conditions ensure stability. The first condition is automatically satisfied because the trace of the matrix is less than zero. The second condition requires that the determinant is greater than zero, which can be equivalently expressed as
\be
\label{B7}
\frac{\bar{f}_{R} + \frac{2}{3}\bar{R}\bar{f}_{X}}{3\bar{f}_{RR} + 2\bar{f}_{X} + 3\bar{R}\bar{f}_{RX} + \frac{3}{4}\bar{R}^2\bar{f}_{XX}} > \frac{\bar{R}}{3}.
\ee
Specifically, when $\bar{f}_{X}$, $\bar{f}_{RX}$, and $\bar{f}_{XX}$ are all zero, the stability condition simplifies to $f(R)$ model.

\appendix
\bibliographystyle{unsrt}
\bibliography{DSnonlocal}

@article{SupernovaCosmologyProject:1998vns,
    author = "Perlmutter, S. and others",
    collaboration = "Supernova Cosmology Project",
    title = "{Measurements of $\Omega$ and $\Lambda$ from 42 high redshift supernovae}",
    eprint = "astro-ph/9812133",
    archivePrefix = "arXiv",
    reportNumber = "LBNL-41801, LBL-41801",
    doi = "10.1086/307221",
    journal = "Astrophys. J.",
    volume = "517",
    pages = "565--586",
    year = "1999"
}

@article{Astier_2006,
   title={The Supernova Legacy Survey: measurement of $\Omega_M$, $\Omega_{\Lambda}$ andwfrom the first year data set},
   volume={447},
   ISSN={1432-0746},
   url={http://dx.doi.org/10.1051/0004-6361:20054185},
   DOI={10.1051/0004-6361:20054185},
   number={1},
   journal={ Astrophysics},
   publisher={EDP Sciences},
   author={Astier, P. and Guy, J. and Regnault, N. and Pain, R. and Aubourg, E. and Balam, D. and Basa, S. and Carlberg, R. G. and Fabbro, S. and Fouchez, D. and Hook, I. M. and Howell, D. A. and Lafoux, H. and Neill, J. D. and Palanque-Delabrouille, N. and Perrett, K. and Pritchet, C. J. and Rich, J. and Sullivan, M. and Taillet, R. and Aldering, G. and Antilogus, P. and Arsenijevic, V. and Balland, C. and Baumont, S. and Bronder, J. and Courtois, H. and Ellis, R. S. and Filiol, M. and Gonçalves, A. C. and Goobar, A. and Guide, D. and Hardin, D. and Lusset, V. and Lidman, C. and McMahon, R. and Mouchet, M. and Mourao, A. and Perlmutter, S. and Ripoche, P. and Tao, C. and Walton, N.},
   year={2006},
   month=jan, pages={31–48} }

@article{SupernovaSearchTeam:1998fmf,
    author = "Riess, Adam G. and others",
    collaboration = "Supernova Search Team",
    title = "{Observational evidence from supernovae for an accelerating universe and a cosmological constant}",
    eprint = "astro-ph/9805201",
    archivePrefix = "arXiv",
    doi = "10.1086/300499",
    journal = "Astron. J.",
    volume = "116",
    pages = "1009--1038",
    year = "1998"
}

@article{SDSS:2003eyi,
    author = "Tegmark, Max and others",
    collaboration = "SDSS",
    title = "{Cosmological parameters from SDSS and WMAP}",
    eprint = "astro-ph/0310723",
    archivePrefix = "arXiv",
    reportNumber = "FERMILAB-PUB-03-435-A",
    doi = "10.1103/PhysRevD.69.103501",
    journal = "Phys. Rev. D",
    volume = "69",
    pages = "103501",
    year = "2004"
}

@article{WMAP:2010qai,
    author = "Komatsu, E. and others",
    collaboration = "WMAP",
    title = "{Seven-Year Wilkinson Microwave Anisotropy Probe (WMAP) Observations: Cosmological Interpretation}",
    eprint = "1001.4538",
    archivePrefix = "arXiv",
    primaryClass = "astro-ph.CO",
    doi = "10.1088/0067-0049/192/2/18",
    journal = "Astrophys. J. Suppl.",
    volume = "192",
    pages = "18",
    year = "2011"
}

@article{SDSS:2004kqt,
    author = "Seljak, Uros and others",
    collaboration = "SDSS",
    title = "{Cosmological parameter analysis including SDSS Ly-alpha forest and galaxy bias: Constraints on the primordial spectrum of fluctuations, neutrino mass, and dark energy}",
    eprint = "astro-ph/0407372",
    archivePrefix = "arXiv",
    reportNumber = "FERMILAB-PUB-04-420-CD",
    doi = "10.1103/PhysRevD.71.103515",
    journal = "Phys. Rev. D",
    volume = "71",
    pages = "103515",
    year = "2005"
}

@article{SDSS:2005xqv,
    author = "Eisenstein, Daniel J. and others",
    collaboration = "SDSS",
    title = "{Detection of the Baryon Acoustic Peak in the Large-Scale Correlation Function of SDSS Luminous Red Galaxies}",
    eprint = "astro-ph/0501171",
    archivePrefix = "arXiv",
    reportNumber = "FERMILAB-PUB-05-057-A-CD",
    doi = "10.1086/466512",
    journal = "Astrophys. J.",
    volume = "633",
    pages = "560--574",
    year = "2005"
}

@article{Jain:2003tba,
    author = "Jain, Bhuvnesh and Taylor, Andy",
    title = "{Cross-correlation tomography: measuring dark energy evolution with weak lensing}",
    eprint = "astro-ph/0306046",
    archivePrefix = "arXiv",
    doi = "10.1103/PhysRevLett.91.141302",
    journal = "Phys. Rev. Lett.",
    volume = "91",
    pages = "141302",
    year = "2003"
}

@article{Kilbinger:2008gk,
    author = "Kilbinger, M. and others",
    title = "{Dark energy constraints and correlations with systematics from CFHTLS weak lensing, SNLS supernovae Ia and WMAP5}",
    eprint = "0810.5129",
    archivePrefix = "arXiv",
    primaryClass = "astro-ph",
    doi = "10.1051/0004-6361/200811247",
    journal = "Astron. Astrophys.",
    volume = "497",
    pages = "677",
    year = "2009"
}

@inproceedings{t1974one,
  title={One-loop divergencies in the theory of gravitation},
  author={t Hooft, Gerard and Veltman, MJG1974AnIHP},
  booktitle={Annales de l'IHP Physique th{\'e}orique},
  volume={20},
  number={1},
  pages={69--94},
  year={1974}
}

@article{Deser:1974cz,
    author = "Deser, Stanley and van Nieuwenhuizen, P.",
    title = "{One Loop Divergences of Quantized Einstein-Maxwell Fields}",
    reportNumber = "Print-74-0576 (BRANDEIS)",
    doi = "10.1103/PhysRevD.10.401",
    journal = "Phys. Rev. D",
    volume = "10",
    pages = "401",
    year = "1974"
}

@article{Deser:1974xq,
    author = "Deser, Stanley and Tsao, Hung-Sheng and van Nieuwenhuizen, P.",
    title = "{One Loop Divergences of the Einstein Yang-Mills System}",
    reportNumber = "Print-74-1164 (BRANDEIS)",
    doi = "10.1103/PhysRevD.10.3337",
    journal = "Phys. Rev. D",
    volume = "10",
    pages = "3337",
    year = "1974"
}

@book{Buchbinder:1992rb,
    author = "Buchbinder, I. L. and Odintsov, S. D. and Shapiro, I. L.",
    title = "{Effective action in quantum gravity}",
    year = "1992"
}

@article{Codello:2007bd,
    author = "Codello, Alessandro and Percacci, Roberto and Rahmede, Christoph",
    title = "{Ultraviolet properties of f(R)-gravity}",
    eprint = "0705.1769",
    archivePrefix = "arXiv",
    primaryClass = "hep-th",
    doi = "10.1142/S0217751X08038135",
    journal = "Int. J. Mod. Phys. A",
    volume = "23",
    pages = "143--150",
    year = "2008"
}

@article{Machado:2007ea,
    author = "Machado, Pedro F. and Saueressig, Frank",
    title = "{On the renormalization group flow of f(R)-gravity}",
    eprint = "0712.0445",
    archivePrefix = "arXiv",
    primaryClass = "hep-th",
    reportNumber = "ITP-UU-07-63, SPIN-07-48, SPHT-T07-154",
    doi = "10.1103/PhysRevD.77.124045",
    journal = "Phys. Rev. D",
    volume = "77",
    pages = "124045",
    year = "2008"
}

@article{Codello:2008vh,
    author = "Codello, Alessandro and Percacci, Roberto and Rahmede, Christoph",
    title = "{Investigating the Ultraviolet Properties of Gravity with a Wilsonian Renormalization Group Equation}",
    eprint = "0805.2909",
    archivePrefix = "arXiv",
    primaryClass = "hep-th",
    doi = "10.1016/j.aop.2008.08.008",
    journal = "Annals Phys.",
    volume = "324",
    pages = "414--469",
    year = "2009"
}

@article{Knorr:2019atm,
    author = "Knorr, Benjamin and Ripken, Chris and Saueressig, Frank",
    title = "{Form Factors in Asymptotic Safety: conceptual ideas and computational toolbox}",
    eprint = "1907.02903",
    archivePrefix = "arXiv",
    primaryClass = "hep-th",
    doi = "10.1088/1361-6382/ab4a53",
    journal = "Class. Quant. Grav.",
    volume = "36",
    number = "23",
    pages = "234001",
    year = "2019"
}

@article{Cognola:2005de,
    author = "Cognola, Guido and Elizalde, Emilio and Nojiri, Shin'ichi and Odintsov, Sergei D. and Zerbini, Sergio",
    title = "{One-loop f(R) gravity in de Sitter universe}",
    eprint = "hep-th/0501096",
    archivePrefix = "arXiv",
    doi = "10.1088/1475-7516/2005/02/010",
    journal = "JCAP",
    volume = "02",
    pages = "010",
    year = "2005"
}

@article{Ruf:2017bqx,
    author = "Ruf, Michael S. and Steinwachs, Christian F.",
    title = "{One-loop divergences for $f(R)$ gravity}",
    eprint = "1711.04785",
    archivePrefix = "arXiv",
    primaryClass = "gr-qc",
    reportNumber = "FR-PHENO-2017-020",
    doi = "10.1103/PhysRevD.97.044049",
    journal = "Phys. Rev. D",
    volume = "97",
    number = "4",
    pages = "044049",
    year = "2018"
}

@article{Alvarez-Gaume:2015rwa,
    author = {Alvarez-Gaume, Luis and Kehagias, Alex and Kounnas, Costas and L\"ust, Dieter and Riotto, Antonio},
    title = "{Aspects of Quadratic Gravity}",
    eprint = "1505.07657",
    archivePrefix = "arXiv",
    primaryClass = "hep-th",
    reportNumber = "LMU-ASC-26-15, MPP-2015-95, CERN-PH-TH-2015-099",
    doi = "10.1002/prop.201500100",
    journal = "Fortsch. Phys.",
    volume = "64",
    number = "2-3",
    pages = "176--189",
    year = "2016"
}

@article{Salam:1978fd,
    author = "Salam, Abdus and Strathdee, J. A.",
    title = "{Remarks on High-energy Stability and Renormalizability of Gravity Theory}",
    reportNumber = "IC/78/12",
    doi = "10.1103/PhysRevD.18.4480",
    journal = "Phys. Rev. D",
    volume = "18",
    pages = "4480",
    year = "1978"
}

@article{Masuda:1976qg,
    author = "Masuda, Naohiko and Weiner, Richard M.",
    title = "{Energy Distribution of Secondaries in Proton-Nucleus Collisions at Very High-Energies}",
    reportNumber = "Print-76-0860 (MARBURG)",
    doi = "10.1016/0370-2693(77)90349-5",
    journal = "Phys. Lett. B",
    volume = "70",
    pages = "77--82",
    year = "1977"
}

@article{Tomboulis:1983sw,
    author = "Tomboulis, E. T.",
    title = "{Unitarity in Higher Derivative Quantum Gravity}",
    reportNumber = "UCLA/83/TEP/18",
    doi = "10.1103/PhysRevLett.52.1173",
    journal = "Phys. Rev. Lett.",
    volume = "52",
    pages = "1173",
    year = "1984"
}

@article{PhysRevD.16.953,
  title = {Renormalization of higher-derivative quantum gravity},
  author = {Stelle, K. S.},
  journal = {Phys. Rev. D},
  volume = {16},
  issue = {4},
  pages = {953--969},
  numpages = {0},
  year = {1977},
  month = {Aug},
  publisher = {American Physical Society},
  doi = {10.1103/PhysRevD.16.953},
  url = {https://link.aps.org/doi/10.1103/PhysRevD.16.953}
}

@article{Tomboulis:1977jk,
    author = "Tomboulis, E.",
    title = "{1/N Expansion and Renormalization in Quantum Gravity}",
    reportNumber = "Print-77-0666 (PRINCETON)",
    doi = "10.1016/0370-2693(77)90678-5",
    journal = "Phys. Lett. B",
    volume = "70",
    pages = "361--364",
    year = "1977"
}

@article{Tomboulis:1980bs,
    author = "Tomboulis, E.",
    title = "{Renormalizability and Asymptotic Freedom in Quantum Gravity}",
    reportNumber = "Print-80-0537 (PRINCETON)",
    doi = "10.1016/0370-2693(80)90550-X",
    journal = "Phys. Lett. B",
    volume = "97",
    pages = "77--80",
    year = "1980"
}

@article{Antoniadis:1986tu,
    author = "Antoniadis, Ignatios and Tomboulis, E. T.",
    title = "{Gauge Invariance and Unitarity in Higher Derivative Quantum Gravity}",
    reportNumber = "UCLA/84/TEP/10",
    doi = "10.1103/PhysRevD.33.2756",
    journal = "Phys. Rev. D",
    volume = "33",
    pages = "2756",
    year = "1986"
}

@article{Deser:2007jk,
    author = "Deser, Stanley and Woodard, R. P.",
    title = "{Nonlocal Cosmology}",
    eprint = "0706.2151",
    archivePrefix = "arXiv",
    primaryClass = "astro-ph",
    reportNumber = "UFIFT-QG-07-03, BRX-TH-589",
    doi = "10.1103/PhysRevLett.99.111301",
    journal = "Phys. Rev. Lett.",
    volume = "99",
    pages = "111301",
    year = "2007"
}

@article{Fradkin:1983mq,
    author = "Fradkin, E. S. and Tseytlin, Arkady A.",
    title = "{One Loop Effective Potential in Gauged O(4) Supergravity}",
    reportNumber = "LEBEDEV-83-135",
    doi = "10.1016/0550-3213(84)90074-9",
    journal = "Nucl. Phys. B",
    volume = "234",
    pages = "472",
    year = "1984"
}

@article{NOJIRI2008821,
title = {Modified non-local-F(R) gravity as the key for the inflation and dark energy},
journal = {Physics Letters B},
volume = {659},
number = {4},
pages = {821-826},
year = {2008},
issn = {0370-2693},
doi = {https://doi.org/10.1016/j.physletb.2007.12.001},
url = {https://www.sciencedirect.com/science/article/pii/S0370269307014876},
author = {Shin'ichi Nojiri and Sergei D. Odintsov}
}

@article{Bamba:2012ky,
    author = "Bamba, Kazuharu and Nojiri, Shinichi and Odintsov, Sergei D. and Sasaki, Misao",
    title = "{Screening of cosmological constant for De Sitter Universe in non-local gravity, phantom-divide crossing and finite-time future singularities}",
    eprint = "1104.2692",
    archivePrefix = "arXiv",
    primaryClass = "hep-th",
    reportNumber = "YITP-11-46",
    doi = "10.1007/s10714-012-1342-7",
    journal = "Gen. Rel. Grav.",
    volume = "44",
    pages = "1321--1356",
    year = "2012"
}

@article{Zhang:2011uv,
    author = "Zhang, Ying-li and Sasaki, Misao",
    title = "{Screening of cosmological constant in non-local cosmology}",
    eprint = "1108.2112",
    archivePrefix = "arXiv",
    primaryClass = "gr-qc",
    reportNumber = "YITP-11-72",
    doi = "10.1142/S021827181250006X",
    journal = "Int. J. Mod. Phys. D",
    volume = "21",
    pages = "1250006",
    year = "2012"
}

@article{Elizalde:2011su,
    author = "Elizalde, E. and Pozdeeva, E. O. and Vernov, S. Yu.",
    title = "{De Sitter Universe in Non-local Gravity}",
    eprint = "1110.5806",
    archivePrefix = "arXiv",
    primaryClass = "astro-ph.CO",
    doi = "10.1103/PhysRevD.85.044002",
    journal = "Phys. Rev. D",
    volume = "85",
    pages = "044002",
    year = "2012"
}

@article{Nojiri:2007uq,
    author = "Nojiri, Shin'ichi and Odintsov, Sergei D.",
    title = "{Modified non-local-F(R) gravity as the key for the inflation and dark energy}",
    eprint = "0708.0924",
    archivePrefix = "arXiv",
    primaryClass = "hep-th",
    reportNumber = "YITP-07-49",
    doi = "10.1016/j.physletb.2007.12.001",
    journal = "Phys. Lett. B",
    volume = "659",
    pages = "821--826",
    year = "2008"
}

@article{Jhingan:2008ym,
    author = "Jhingan, S. and Nojiri, S. and Odintsov, S. D. and Sami, M. and Thongkool, I and Zerbini, S.",
    title = "{Phantom and non-phantom dark energy: The Cosmological relevance of non-locally corrected gravity}",
    eprint = "0803.2613",
    archivePrefix = "arXiv",
    primaryClass = "hep-th",
    doi = "10.1016/j.physletb.2008.04.054",
    journal = "Phys. Lett. B",
    volume = "663",
    pages = "424--428",
    year = "2008"
}

@article{Koshelev:2016xqb,
    author = "Koshelev, Alexey S. and Modesto, Leonardo and Rachwal, Leslaw and Starobinsky, Alexei A.",
    title = "{Occurrence of exact $R^2$ inflation in non-local UV-complete gravity}",
    eprint = "1604.03127",
    archivePrefix = "arXiv",
    primaryClass = "hep-th",
    doi = "10.1007/JHEP11(2016)067",
    journal = "JHEP",
    volume = "11",
    pages = "067",
    year = "2016"
}

@article{Modesto:2011kw,
    author = "Modesto, Leonardo",
    title = "{Super-renormalizable Quantum Gravity}",
    eprint = "1107.2403",
    archivePrefix = "arXiv",
    primaryClass = "hep-th",
    doi = "10.1103/PhysRevD.86.044005",
    journal = "Phys. Rev. D",
    volume = "86",
    pages = "044005",
    year = "2012"
}

@article{Modesto:2017sdr,
    author = "Modesto, Leonardo and Rachwa\l{}, Les\l{}aw",
    title = "{Nonlocal quantum gravity: A review}",
    doi = "10.1142/S0218271817300208",
    journal = "Int. J. Mod. Phys. D",
    volume = "26",
    number = "11",
    pages = "1730020",
    year = "2017"
}

@article{Belgacem:2017cqo,
    author = "Belgacem, Enis and Dirian, Yves and Foffa, Stefano and Maggiore, Michele",
    title = "{Nonlocal gravity. Conceptual aspects and cosmological predictions}",
    eprint = "1712.07066",
    archivePrefix = "arXiv",
    primaryClass = "hep-th",
    doi = "10.1088/1475-7516/2018/03/002",
    journal = "JCAP",
    volume = "03",
    pages = "002",
    year = "2018"
}

@article{Nojiri:2010pw,
    author = "Nojiri, Shin'ichi and Odintsov, Sergei D. and Sasaki, Misao and Zhang, Ying-li",
    title = "{Screening of cosmological constant in non-local gravity}",
    eprint = "1010.5375",
    archivePrefix = "arXiv",
    primaryClass = "gr-qc",
    reportNumber = "YITP-10-91",
    doi = "10.1016/j.physletb.2010.12.035",
    journal = "Phys. Lett. B",
    volume = "696",
    pages = "278--282",
    year = "2011"
}

@article{Zhang:2016ykx,
    author = "Zhang, Ying-li and Koyama, Kazuya and Sasaki, Misao and Zhao, Gong-Bo",
    title = "{Acausality in Nonlocal Gravity Theory}",
    eprint = "1601.03808",
    archivePrefix = "arXiv",
    primaryClass = "hep-th",
    reportNumber = "YITP-16-3",
    doi = "10.1007/JHEP03(2016)039",
    journal = "JHEP",
    volume = "03",
    pages = "039",
    year = "2016"
}

@article{Nojiri:2017ncd,
    author = "Nojiri, S. and Odintsov, S. D. and Oikonomou, V. K.",
    title = "{Modified Gravity Theories on a Nutshell: Inflation, Bounce and Late-time Evolution}",
    eprint = "1705.11098",
    archivePrefix = "arXiv",
    primaryClass = "gr-qc",
    reportNumber = "PHYS.REPT.-692-(2017)-1-104, Phys.Rept. 692 (2017) 1-104",
    doi = "10.1016/j.physrep.2017.06.001",
    journal = "Phys. Rept.",
    volume = "692",
    pages = "1--104",
    year = "2017"
}

@article{Nojiri:2010wj,
    author = "Nojiri, Shin'ichi and Odintsov, Sergei D.",
    title = "{Unified cosmic history in modified gravity: from F(R) theory to Lorentz non-invariant models}",
    eprint = "1011.0544",
    archivePrefix = "arXiv",
    primaryClass = "gr-qc",
    doi = "10.1016/j.physrep.2011.04.001",
    journal = "Phys. Rept.",
    volume = "505",
    pages = "59--144",
    year = "2011"
}

@article{Nojiri:2006ri,
    author = "Nojiri, Shin'ichi and Odintsov, Sergei D.",
    editor = "Borowiec, Andrzej",
    title = "{Introduction to modified gravity and gravitational alternative for dark energy}",
    eprint = "hep-th/0601213",
    archivePrefix = "arXiv",
    reportNumber = "KARP-2006-06",
    doi = "10.1142/S0219887807001928",
    journal = "eConf",
    volume = "C0602061",
    pages = "06",
    year = "2006"
}

@article{Capozziello:2011et,
    author = "Capozziello, Salvatore and De Laurentis, Mariafelicia",
    title = "{Extended Theories of Gravity}",
    eprint = "1108.6266",
    archivePrefix = "arXiv",
    primaryClass = "gr-qc",
    doi = "10.1016/j.physrep.2011.09.003",
    journal = "Phys. Rept.",
    volume = "509",
    pages = "167--321",
    year = "2011"
}

@book{Faraoni:2010pgm,
    author = "Faraoni, Valerio and Capozziello, Salvatore",
    title = "{Beyond Einstein Gravity}: {A Survey of Gravitational Theories for Cosmology and Astrophysics}",
    doi = "10.1007/978-94-007-0165-6",
    isbn = "978-94-007-0164-9, 978-94-007-0165-6",
    publisher = "Springer",
    address = "Dordrecht",
    year = "2011"
}

@article{delaCruz-Dombriz:2012bni,
    author = "de la Cruz-Dombriz, A. and Saez-Gomez, D.",
    title = "{Black holes, cosmological solutions, future singularities, and their thermodynamical properties in modified gravity theories}",
    eprint = "1207.2663",
    archivePrefix = "arXiv",
    primaryClass = "gr-qc",
    doi = "10.3390/e14091717",
    journal = "Entropy",
    volume = "14",
    pages = "1717--1770",
    year = "2012"
}

@article{Olmo:2011uz,
    author = "Olmo, Gonzalo J.",
    title = "{Palatini Approach to Modified Gravity: f(R) Theories and Beyond}",
    eprint = "1101.3864",
    archivePrefix = "arXiv",
    primaryClass = "gr-qc",
    doi = "10.1142/S0218271811018925",
    journal = "Int. J. Mod. Phys. D",
    volume = "20",
    pages = "413--462",
    year = "2011"
}

@article{PhysRevLett.116.031302,
  title = {Improved Constraints on Cosmology and Foregrounds from BICEP2 and Keck Array Cosmic Microwave Background Data with Inclusion of 95 GHz Band},
  author = {Ade, P. A. R. and Ahmed, Z. and Aikin, R. W. and Alexander, K. D. and Barkats, D. and Benton, S. J. and Bischoff, C. A. and Bock, J. J. and Bowens-Rubin, R. and Brevik, J. A. and Buder, I. and Bullock, E. and Buza, V. and Connors, J. and Crill, B. P. and Duband, L. and Dvorkin, C. and Filippini, J. P. and Fliescher, S. and Grayson, J. and Halpern, M. and Harrison, S. and Hilton, G. C. and Hui, H. and Irwin, K. D. and Karkare, K. S. and Karpel, E. and Kaufman, J. P. and Keating, B. G. and Kefeli, S. and Kernasovskiy, S. A. and Kovac, J. M. and Kuo, C. L. and Leitch, E. M. and Lueker, M. and Megerian, K. G. and Netterfield, C. B. and Nguyen, H. T. and O'Brient, R. and Ogburn, R. W. and Orlando, A. and Pryke, C. and Richter, S. and Schwarz, R. and Sheehy, C. D. and Staniszewski, Z. K. and Steinbach, B. and Sudiwala, R. V. and Teply, G. P. and Thompson, K. L. and Tolan, J. E. and Tucker, C. and Turner, A. D. and Vieregg, A. G. and Weber, A. C. and Wiebe, D. V. and Willmert, J. and Wong, C. L. and Wu, W. L. K. and Yoon, K. W.},
  collaboration = {Keck Array and BICEP2 Collaborations},
  journal = {Phys. Rev. Lett.},
  volume = {116},
  issue = {3},
  pages = {031302},
  numpages = {9},
  year = {2016},
  month = {Jan},
  publisher = {American Physical Society},
  doi = {10.1103/PhysRevLett.116.031302},
  url = {https://link.aps.org/doi/10.1103/PhysRevLett.116.031302}
}

@article{Koshelev_2016,
   title={Occurrence of exact R 2 inflation in non-local UV-complete gravity},
   volume={2016},
   ISSN={1029-8479},
   url={http://dx.doi.org/10.1007/JHEP11(2016)067},
   DOI={10.1007/jhep11(2016)067},
   number={11},
   journal={Journal of High Energy Physics},
   publisher={Springer Science and Business Media LLC},
   author={Koshelev, Alexey S. and Modesto, Leonardo and Rachwal, Leslaw and Starobinsky, Alexei A.},
   year={2016},
   month=nov }

@article{Planck:2015fie,
    author = "Ade, P. A. R. and others",
    collaboration = "Planck",
    title = "{Planck 2015 results. XIII. Cosmological parameters}",
    eprint = "1502.01589",
    archivePrefix = "arXiv",
    primaryClass = "astro-ph.CO",
    doi = "10.1051/0004-6361/201525830",
    journal = "Astron. Astrophys.",
    volume = "594",
    pages = "A13",
    year = "2016"
}

@article{Planck:2015sxf,
    author = "Ade, P. A. R. and others",
    collaboration = "Planck",
    title = "{Planck 2015 results. XX. Constraints on inflation}",
    eprint = "1502.02114",
    archivePrefix = "arXiv",
    primaryClass = "astro-ph.CO",
    doi = "10.1051/0004-6361/201525898",
    journal = "Astron. Astrophys.",
    volume = "594",
    pages = "A20",
    year = "2016"
}

@misc{tomboulis1997superrenormalizablegaugegravitationaltheories,
      title={Superrenormalizable gauge and gravitational theories}, 
      author={E. T. Tomboulis},
      year={1997},
      eprint={hep-th/9702146},
      archivePrefix={arXiv},
      primaryClass={hep-th},
      url={https://arxiv.org/abs/hep-th/9702146}, 
}

@misc{delhom2022instabilitiesfieldtheorieslecture,
      title={Instabilities in field theories: Lecture notes with a view into modified gravity}, 
      author={Adrià Delhom and Alejandro Jiménez-Cano and Francisco José Maldonado Torralba},
      year={2022},
      eprint={2207.13431},
      archivePrefix={arXiv},
      primaryClass={gr-qc},
      url={https://arxiv.org/abs/2207.13431}, 
}

@article{Arbuzova:2019xti,
    author = "Arbuzova, E. V. and Dolgov, A. D.",
    title = "{Instability Effects in $F(R)$-Modified Gravity and in Gravitational Baryogenesis}",
    doi = "10.1134/S1063779619060078",
    journal = "Phys. Part. Nucl.",
    volume = "50",
    number = "6",
    pages = "850--943",
    year = "2019"
}

@ARTICLE{2014PhLB..739..279A,
       author = {{Arbuzova}, E.~V. and {Dolgov}, A.~D. and {Reverberi}, L.},
        title = "{Jeans instability in classical and modified gravity}",
      journal = {Physics Letters B},
         year = 2014,
        month = dec,
       volume = {739},
        pages = {279-284},
          doi = {10.1016/j.physletb.2014.11.004},
archivePrefix = {arXiv},
       eprint = {1406.7104},
 primaryClass = {gr-qc},
       adsurl = {https://ui.adsabs.harvard.edu/abs/2014PhLB..739..279A}
}

@article{Felice_2006,
   title={Ghosts, instabilities, and superluminal propagation in modified gravity models},
   volume={2006},
   ISSN={1475-7516},
   url={http://dx.doi.org/10.1088/1475-7516/2006/08/005},
   DOI={10.1088/1475-7516/2006/08/005},
   number={08},
   journal={Journal of Cosmology and Astroparticle Physics},
   publisher={IOP Publishing},
   author={Felice, Antonio De and Hindmarsh, Mark and Trodden, Mark},
   year={2006},
   month=aug, pages={005–005} }

@article{DeFelice:2006pg,
    author = "De Felice, Antonio and Hindmarsh, Mark and Trodden, Mark",
    title = "{Ghosts, Instabilities, and Superluminal Propagation in Modified Gravity Models}",
    eprint = "astro-ph/0604154",
    archivePrefix = "arXiv",
    doi = "10.1088/1475-7516/2006/08/005",
    journal = "JCAP",
    volume = "08",
    pages = "005",
    year = "2006"
}

@article{PhysRevD.109.023515,
  title = {Inflation and late-time accelerated expansion driven by $k$-essence degenerate dynamics},
  author = {Ferreira, Alexsandre L. and Pinto-Neto, Nelson and Zanelli, Jorge},
  journal = {Phys. Rev. D},
  volume = {109},
  issue = {2},
  pages = {023515},
  numpages = {12},
  year = {2024},
  month = {Jan},
  publisher = {American Physical Society},
  doi = {10.1103/PhysRevD.109.023515},
  url = {https://link.aps.org/doi/10.1103/PhysRevD.109.023515}
}

@article{Cicoli:2023opf,
    author = "Cicoli, Michele and Conlon, Joseph P. and Maharana, Anshuman and Parameswaran, Susha and Quevedo, Fernando and Zavala, Ivonne",
    title = "{String cosmology: From the early universe to today}",
    eprint = "2303.04819",
    archivePrefix = "arXiv",
    primaryClass = "hep-th",
    doi = "10.1016/j.physrep.2024.01.002",
    journal = "Phys. Rept.",
    volume = "1059",
    pages = "1--155",
    year = "2024"
}

@article{Vazquez:2018qdg,
    author = "V\'azquez, J. Alberto and Padilla, Luis E. and Matos, Tonatiuh",
    title = "{Inflationary cosmology: from theory to observations}",
    eprint = "1810.09934",
    archivePrefix = "arXiv",
    primaryClass = "astro-ph.CO",
    doi = "10.31349/RevMexFisE.17.73",
    journal = "Rev. Mex. Fis. E",
    volume = "17",
    number = "1",
    pages = "73--91",
    year = "2020"
}

@article{RevModPhys.61.1,
  title = {The cosmological constant problem},
  author = {Weinberg, Steven},
  journal = {Rev. Mod. Phys.},
  volume = {61},
  issue = {1},
  pages = {1--23},
  numpages = {0},
  year = {1989},
  month = {Jan},
  publisher = {American Physical Society},
  doi = {10.1103/RevModPhys.61.1},
  url = {https://link.aps.org/doi/10.1103/RevModPhys.61.1}
}

@article{Sol_Peracaula_2022,
   title={The cosmological constant problem and running vacuum in the expanding universe},
   volume={380},
   ISSN={1471-2962},
   url={http://dx.doi.org/10.1098/rsta.2021.0182},
   DOI={10.1098/rsta.2021.0182},
   number={2230},
   journal={Philosophical Transactions of the Royal Society A: Mathematical, Physical and Engineering Sciences},
   publisher={The Royal Society},
   author={Solà Peracaula, Joan},
   year={2022},
   month=jul }

@article{PhysRevD.106.083510,
  title = {Cosmological constant problem on the horizon},
  author = {Firouzjahi, Hassan},
  journal = {Phys. Rev. D},
  volume = {106},
  issue = {8},
  pages = {083510},
  numpages = {18},
  year = {2022},
  month = {Oct},
  publisher = {American Physical Society},
  doi = {10.1103/PhysRevD.106.083510},
  url = {https://link.aps.org/doi/10.1103/PhysRevD.106.083510}
}

@article{Koshelev:2013lfm,
    author = "Koshelev, Alexey S.",
    title = "{Stable analytic bounce in non-local Einstein-Gauss-Bonnet cosmology}",
    eprint = "1302.2140",
    archivePrefix = "arXiv",
    primaryClass = "astro-ph.CO",
    doi = "10.1088/0264-9381/30/15/155001",
    journal = "Class. Quant. Grav.",
    volume = "30",
    pages = "155001",
    year = "2013"
}

@article{Dimitrijevic:2015eaa,
    author = "Dimitrijevi\'c, Ivan and Dragovich, Branko and Grujic, J. and Koshelev, Alexey S. and Raki\'c, Zoran and Stankovi\'c, Jelena",
    title = "{Cosmology of non-local f(R) gravity}",
    eprint = "1509.04254",
    archivePrefix = "arXiv",
    primaryClass = "hep-th",
    doi = "10.2298/FIL1904163D",
    journal = "Filomat",
    volume = "33",
    number = "4",
    pages = "1163--1178",
    year = "2019"
}

@article{Briscese:2019rii,
    author = "Briscese, Fabio and Calcagni, Gianluca and Modesto, Leonardo",
    title = "{Nonlinear stability in nonlocal gravity}",
    eprint = "1901.03267",
    archivePrefix = "arXiv",
    primaryClass = "gr-qc",
    doi = "10.1103/PhysRevD.99.084041",
    journal = "Phys. Rev. D",
    volume = "99",
    number = "8",
    pages = "084041",
    year = "2019"
}

@article{Briscese:2018bny,
    author = "Briscese, Fabio and Modesto, Leonardo",
    title = "{Nonlinear stability of Minkowski spacetime in Nonlocal Gravity}",
    eprint = "1811.05117",
    archivePrefix = "arXiv",
    primaryClass = "gr-qc",
    doi = "10.1088/1475-7516/2019/07/009",
    journal = "JCAP",
    volume = "07",
    pages = "009",
    year = "2019"
}

@article{Calcagni_2018,
   title={Black-hole stability in non-local gravity},
   volume={783},
   ISSN={0370-2693},
   url={http://dx.doi.org/10.1016/j.physletb.2018.06.041},
   DOI={10.1016/j.physletb.2018.06.041},
   journal={Physics Letters B},
   publisher={Elsevier BV},
   author={Calcagni, Gianluca and Modesto, Leonardo and Myung, Yun Soo},
   year={2018},
   month=aug, pages={19–23} }

@article{Calcagni_2017,
   title={Stability of Schwarzschild singularity in non-local gravity},
   volume={773},
   ISSN={0370-2693},
   url={http://dx.doi.org/10.1016/j.physletb.2017.09.018},
   DOI={10.1016/j.physletb.2017.09.018},
   journal={Physics Letters B},
   publisher={Elsevier BV},
   author={Calcagni, Gianluca and Modesto, Leonardo},
   year={2017},
   month=oct, pages={596–600} }

@article{Calcagni_2008,
   title={Localization of nonlocal theories},
   volume={662},
   ISSN={0370-2693},
   url={http://dx.doi.org/10.1016/j.physletb.2008.03.024},
   DOI={10.1016/j.physletb.2008.03.024},
   number={3},
   journal={Physics Letters B},
   publisher={Elsevier BV},
   author={Calcagni, Gianluca and Montobbio, Michele and Nardelli, Giuseppe},
   year={2008},
   month=apr, pages={285–289} }

@article{D_Hoker_1999,
   title={Graviton and gauge boson propagators in AdS+1},
   volume={562},
   ISSN={0550-3213},
   url={http://dx.doi.org/10.1016/S0550-3213(99)00524-6},
   DOI={10.1016/s0550-3213(99)00524-6},
   number={1–2},
   journal={Nuclear Physics B},
   publisher={Elsevier BV},
   author={D’Hoker, Eric and Freedman, Daniel Z. and Mathur, Samir D. and Matusis, Alec and Rastelli, Leonardo},
   year={1999},
   month=nov, pages={330–352} }

@article{Craps:2014wga,
    author = "Craps, Ben and De Jonckheere, Tim and Koshelev, Alexey S.",
    title = "{Cosmological perturbations in non-local higher-derivative gravity}",
    eprint = "1407.4982",
    archivePrefix = "arXiv",
    primaryClass = "hep-th",
    doi = "10.1088/1475-7516/2014/11/022",
    journal = "JCAP",
    volume = "11",
    pages = "022",
    year = "2014"
}

@article{Biswas:2012bp,
    author = "Biswas, Tirthabir and Koshelev, Alexey S. and Mazumdar, Anupam and Vernov, Sergey Yu.",
    title = "{Stable bounce and inflation in non-local higher derivative cosmology}",
    eprint = "1206.6374",
    archivePrefix = "arXiv",
    primaryClass = "astro-ph.CO",
    doi = "10.1088/1475-7516/2012/08/024",
    journal = "JCAP",
    volume = "08",
    pages = "024",
    year = "2012"
}

@article{Biswas:2016egy,
    author = "Biswas, Tirthabir and Koshelev, Alexey S. and Mazumdar, Anupam",
    title = "{Consistent higher derivative gravitational theories with stable de Sitter and anti\textendash{}de Sitter backgrounds}",
    eprint = "1606.01250",
    archivePrefix = "arXiv",
    primaryClass = "gr-qc",
    doi = "10.1103/PhysRevD.95.043533",
    journal = "Phys. Rev. D",
    volume = "95",
    number = "4",
    pages = "043533",
    year = "2017"
}

@article{Biswas_2012,
   title={Towards Singularity- and Ghost-Free Theories of Gravity},
   volume={108},
   ISSN={1079-7114},
   url={http://dx.doi.org/10.1103/PhysRevLett.108.031101},
   DOI={10.1103/physrevlett.108.031101},
   number={3},
   journal={Physical Review Letters},
   publisher={American Physical Society (APS)},
   author={Biswas, Tirthabir and Gerwick, Erik and Koivisto, Tomi and Mazumdar, Anupam},
   year={2012},
   month=jan }

@article{Krasnikov:1987yj,
    author = "Krasnikov, N. V.",
    title = "{NONLOCAL GAUGE THEORIES}",
    doi = "10.1007/BF01017588",
    journal = "Theor. Math. Phys.",
    volume = "73",
    pages = "1184--1190",
    year = "1987"
}

@article{Tomboulis:1997gg,
    author = "Tomboulis, E. T.",
    title = "{Superrenormalizable gauge and gravitational theories}",
    eprint = "hep-th/9702146",
    archivePrefix = "arXiv",
    reportNumber = "UCLA-97-TEP-2",
    month = "2",
    year = "1997"
}

@article{Koshelev:2017ebj,
    author = "Koshelev, Alexey S. and Sravan Kumar, K. and Modesto, Leonardo and Rachwa\l{}, Les\l{}aw",
    title = "{Finite quantum gravity in dS and AdS spacetimes}",
    eprint = "1710.07759",
    archivePrefix = "arXiv",
    primaryClass = "hep-th",
    doi = "10.1103/PhysRevD.98.046007",
    journal = "Phys. Rev. D",
    volume = "98",
    number = "4",
    pages = "046007",
    year = "2018"
}

@article{Biswas:2005qr,
    author = "Biswas, Tirthabir and Mazumdar, Anupam and Siegel, Warren",
    title = "{Bouncing universes in string-inspired gravity}",
    eprint = "hep-th/0508194",
    archivePrefix = "arXiv",
    reportNumber = "YITP-SB-05-23, NORDITA-2005-53",
    doi = "10.1088/1475-7516/2006/03/009",
    journal = "JCAP",
    volume = "03",
    pages = "009",
    year = "2006"
}

@article{Biswas:2010zk,
    author = "Biswas, Tirthabir and Koivisto, Tomi and Mazumdar, Anupam",
    title = "{Towards a resolution of the cosmological singularity in non-local higher derivative theories of gravity}",
    eprint = "1005.0590",
    archivePrefix = "arXiv",
    primaryClass = "hep-th",
    doi = "10.1088/1475-7516/2010/11/008",
    journal = "JCAP",
    volume = "11",
    pages = "008",
    year = "2010"
}

@article{VanNieuwenhuizen:1973fi,
    author = "Van Nieuwenhuizen, P.",
    title = "{On ghost-free tensor lagrangians and linearized gravitation}",
    doi = "10.1016/0550-3213(73)90194-6",
    journal = "Nucl. Phys. B",
    volume = "60",
    pages = "478--492",
    year = "1973"
}

@article{N_ez_2005,
   title={Ghost constraints on modified gravity},
   volume={608},
   ISSN={0370-2693},
   url={http://dx.doi.org/10.1016/j.physletb.2005.01.015},
   DOI={10.1016/j.physletb.2005.01.015},
   number={3–4},
   journal={Physics Letters B},
   publisher={Elsevier BV},
   author={Núñez, Alvaro and Solganik, Slava},
   year={2005},
   month=feb, pages={189–193} }

@article{Chiba_2005,
   title={Generalized gravity and a ghost},
   volume={2005},
   ISSN={1475-7516},
   url={http://dx.doi.org/10.1088/1475-7516/2005/03/008},
   DOI={10.1088/1475-7516/2005/03/008},
   number={03},
   journal={Journal of Cosmology and Astroparticle Physics},
   publisher={IOP Publishing},
   author={Chiba, Takeshi},
   year={2005},}

@article{Feng_2024,
   title={Stability of the de-Sitter universe: one-loop nonlocal f (R) gravity},
   volume={2024},
   ISSN={1029-8479},
   url={http://dx.doi.org/10.1007/JHEP05(2024)115},
   DOI={10.1007/jhep05(2024)115},
   number={5},
   journal={Journal of High Energy Physics},
   publisher={Springer Science and Business Media LLC},
   author={Feng, Haiyuan and Liao, Yi and Yang, Rong-Jia},
   year={2024},
   month=may }

@article{Balasubramanian:2002zh,
    author = "Balasubramanian, Vijay and de Boer, Jan and Minic, Djordje",
    editor = "de Wit, B. and Vandoren, S.",
    title = "{Notes on de Sitter space and holography}",
    eprint = "hep-th/0207245",
    archivePrefix = "arXiv",
    reportNumber = "VPI-IPPAP-02-05, UPR-1008-T, IFTA-2002-26",
    doi = "10.1016/S0003-4916(02)00020-9",
    journal = "Class. Quant. Grav.",
    volume = "19",
    pages = "5655--5700",
    year = "2002"
}

@article{Perlmutter_1999,
   title={Measurements of $\Omega$ and $Lambda$ from 42 High‐Redshift Supernovae},
   volume={517},
   ISSN={1538-4357},
   url={http://dx.doi.org/10.1086/307221},
   DOI={10.1086/307221},
   number={2},
   journal={The Astrophysical Journal},
   publisher={American Astronomical Society},
   author={Perlmutter, S. and Aldering, G. and Goldhaber, G. and Knop, R. A. and Nugent, P. and Castro, P. G. and Deustua, S. and Fabbro, S. and Goobar, A. and Groom, D. E. and Hook, I. M. and Kim, A. G. and Kim, M. Y. and Lee, J. C. and Nunes, N. J. and Pain, R. and Pennypacker, C. R. and Quimby, R. and Lidman, C. and Ellis, R. S. and Irwin, M. and McMahon, R. G. and Ruiz‐Lapuente, P. and Walton, N. and Schaefer, B. and Boyle, B. J. and Filippenko, A. V. and Matheson, T. and Fruchter, A. S. and Panagia, N. and Newberg, H. J. M. and Couch, W. J. and Project, The Supernova Cosmology},
   year={1999},
   month=jun, pages={565–586} }

@article{Riess_1998,
   title={Observational Evidence from Supernovae for an Accelerating Universe and a Cosmological Constant},
   volume={116},
   ISSN={0004-6256},
   url={http://dx.doi.org/10.1086/300499},
   DOI={10.1086/300499},
   number={3},
   journal={The Astronomical Journal},
   publisher={American Astronomical Society},
   author={Riess, Adam G. and Filippenko, Alexei V. and Challis, Peter and Clocchiatti, Alejandro and Diercks, Alan and Garnavich, Peter M. and Gilliland, Ron L. and Hogan, Craig J. and Jha, Saurabh and Kirshner, Robert P. and Leibundgut, B. and Phillips, M. M. and Reiss, David and Schmidt, Brian P. and Schommer, Robert A. and Smith, R. Chris and Spyromilio, J. and Stubbs, Christopher and Suntzeff, Nicholas B. and Tonry, John},
   year={1998},
   month=sep, pages={1009–1038} }

@article{Spergel_2003,
   title={First‐Year
                    Wilkinson Microwave Anisotropy Probe
                    (
                    WMAP
                    ) Observations: Determination of Cosmological Parameters},
   volume={148},
   ISSN={1538-4365},
   url={http://dx.doi.org/10.1086/377226},
   DOI={10.1086/377226},
   number={1},
   journal={The Astrophysical Journal Supplement Series},
   publisher={American Astronomical Society},
   author={Spergel, D. N. and Verde, L. and Peiris, H. V. and Komatsu, E. and Nolta, M. R. and Bennett, C. L. and Halpern, M. and Hinshaw, G. and Jarosik, N. and Kogut, A. and Limon, M. and Meyer, S. S. and Page, L. and Tucker, G. S. and Weiland, J. L. and Wollack, E. and Wright, E. L.},
   year={2003},
   month=sep, pages={175–194} }

@article{Spergel_2007,
   title={Three‐YearWilkinson Microwave Anisotropy Probe(WMAP) Observations: Implications for Cosmology},
   volume={170},
   ISSN={1538-4365},
   url={http://dx.doi.org/10.1086/513700},
   DOI={10.1086/513700},
   number={2},
   journal={The Astrophysical Journal Supplement Series},
   publisher={American Astronomical Society},
   author={Spergel, D. N. and Bean, R. and Dore, O. and Nolta, M. R. and Bennett, C. L. and Dunkley, J. and Hinshaw, G. and Jarosik, N. and Komatsu, E. and Page, L. and Peiris, H. V. and Verde, L. and Halpern, M. and Hill, R. S. and Kogut, A. and Limon, M. and Meyer, S. S. and Odegard, N. and Tucker, G. S. and Weiland, J. L. and Wollack, E. and Wright, E. L.},
   year={2007},
   month=jun, pages={377–408} }

@article{Komatsu_2009,
   title={FIVE-YEARWILKINSON MICROWAVE ANISOTROPY PROBEOBSERVATIONS: COSMOLOGICAL INTERPRETATION},
   volume={180},
   ISSN={1538-4365},
   url={http://dx.doi.org/10.1088/0067-0049/180/2/330},
   DOI={10.1088/0067-0049/180/2/330},
   number={2},
   journal={The Astrophysical Journal Supplement Series},
   publisher={American Astronomical Society},
   author={Komatsu, E. and Dunkley, J. and Nolta, M. R. and Bennett, C. L. and Gold, B. and Hinshaw, G. and Jarosik, N. and Larson, D. and Limon, M. and Page, L. and Spergel, D. N. and Halpern, M. and Hill, R. S. and Kogut, A. and Meyer, S. S. and Tucker, G. S. and Weiland, J. L. and Wollack, E. and Wright, E. L.},
   year={2009},
   month=feb, pages={330–376} }

@article{Komatsu_2011,
   title={SEVEN-YEARWILKINSON MICROWAVE ANISOTROPY PROBE(WMAP) OBSERVATIONS: COSMOLOGICAL INTERPRETATION},
   volume={192},
   ISSN={1538-4365},
   url={http://dx.doi.org/10.1088/0067-0049/192/2/18},
   DOI={10.1088/0067-0049/192/2/18},
   number={2},
   journal={The Astrophysical Journal Supplement Series},
   publisher={American Astronomical Society},
   author={Komatsu, E. and Smith, K. M. and Dunkley, J. and Bennett, C. L. and Gold, B. and Hinshaw, G. and Jarosik, N. and Larson, D. and Nolta, M. R. and Page, L. and Spergel, D. N. and Halpern, M. and Hill, R. S. and Kogut, A. and Limon, M. and Meyer, S. S. and Odegard, N. and Tucker, G. S. and Weiland, J. L. and Wollack, E. and Wright, E. L.},
   year={2011},
   month=jan, pages={18} }

@article{Eisenstein_2005,
   title={Detection of the Baryon Acoustic Peak in the Large‐Scale Correlation Function of SDSS Luminous Red Galaxies},
   volume={633},
   ISSN={1538-4357},
   url={http://dx.doi.org/10.1086/466512},
   DOI={10.1086/466512},
   number={2},
   journal={The Astrophysical Journal},
   publisher={American Astronomical Society},
   author={Eisenstein, Daniel J. and Zehavi, Idit and Hogg, David W. and Scoccimarro, Roman and Blanton, Michael R. and Nichol, Robert C. and Scranton, Ryan and Seo, Hee‐Jong and Tegmark, Max and Zheng, Zheng and Anderson, Scott F. and Annis, Jim and Bahcall, Neta and Brinkmann, Jon and Burles, Scott and Castander, Francisco J. and Connolly, Andrew and Csabai, Istvan and Doi, Mamoru and Fukugita, Masataka and Frieman, Joshua A. and Glazebrook, Karl and Gunn, James E. and Hendry, John S. and Hennessy, Gregory and Ivezić, Zeljko and Kent, Stephen and Knapp, Gillian R. and Lin, Huan and Loh, Yeong‐Shang and Lupton, Robert H. and Margon, Bruce and McKay, Timothy A. and Meiksin, Avery and Munn, Jeffery A. and Pope, Adrian and Richmond, Michael W. and Schlegel, David and Schneider, Donald P. and Shimasaku, Kazuhiro and Stoughton, Christopher and Strauss, Michael A. and SubbaRao, Mark and Szalay, Alexander S. and Szapudi, Istvan and Tucker, Douglas L. and Yanny, Brian and York, Donald G.},
   year={2005},
   month=nov, pages={560–574} }

@article{Jain_2003,
   title={Cross-Correlation Tomography: Measuring Dark Energy Evolution with Weak Lensing},
   volume={91},
   ISSN={1079-7114},
   url={http://dx.doi.org/10.1103/PhysRevLett.91.141302},
   DOI={10.1103/physrevlett.91.141302},
   number={14},
   journal={Physical Review Letters},
   publisher={American Physical Society (APS)},
   author={Jain, Bhuvnesh and Taylor, Andy},
   year={2003},
   month=oct }

@book{Hawking:1973uf,
    author = "Hawking, Stephen W. and Ellis, George F. R.",
    title = "{The Large Scale Structure of Space-Time}",
    doi = "10.1017/9781009253161",
    isbn = "978-1-009-25316-1, 978-1-009-25315-4, 978-0-521-20016-5, 978-0-521-09906-6, 978-0-511-82630-6, 978-0-521-09906-6",
    publisher = "Cambridge University Press",
    series = "Cambridge Monographs on Mathematical Physics",
    month = "2",
    year = "2023"
}

@article{tHooft:1974toh,
    author = "'t Hooft, Gerard and Veltman, M. J. G.",
    title = "{One loop divergencies in the theory of gravitation}",
    journal = "Ann. Inst. H. Poincare A Phys. Theor.",
    volume = "20",
    pages = "69--94",
    year = "1974"
}

@article{Starobinsky:1980te,
    author = "Starobinsky, Alexei A.",
    editor = "Khalatnikov, I. M. and Mineev, V. P.",
    title = "{A New Type of Isotropic Cosmological Models Without Singularity}",
    doi = "10.1016/0370-2693(80)90670-X",
    journal = "Phys. Lett. B",
    volume = "91",
    pages = "99--102",
    year = "1980"
}

@article{Koshelev:2017tvv,
    author = "Koshelev, Alexey S. and Sravan Kumar, K. and Starobinsky, Alexei A.",
    title = "{$R^2$ inflation to probe non-perturbative quantum gravity}",
    eprint = "1711.08864",
    archivePrefix = "arXiv",
    primaryClass = "hep-th",
    doi = "10.1007/JHEP03(2018)071",
    journal = "JHEP",
    volume = "03",
    pages = "071",
    year = "2018"
}

@article{Talaganis_2015,
   title={Towards understanding the ultraviolet behavior of quantum loops in infinite-derivative theories of gravity},
   volume={32},
   ISSN={1361-6382},
   url={http://dx.doi.org/10.1088/0264-9381/32/21/215017},
   DOI={10.1088/0264-9381/32/21/215017},
   number={21},
   journal={Classical and Quantum Gravity},
   publisher={IOP Publishing},
   author={Talaganis, Spyridon and Biswas, Tirthabir and Mazumdar, Anupam},
   year={2015},
   month=oct, pages={215017} }

@article{Don__2015,
   title={Scattering amplitudes in super-renormalizable gravity},
   volume={2015},
   ISSN={1029-8479},
   url={http://dx.doi.org/10.1007/JHEP08(2015)038},
   DOI={10.1007/jhep08(2015)038},
   number={8},
   journal={Journal of High Energy Physics},
   publisher={Springer Science and Business Media LLC},
   author={Donà, Pietro and Giaccari, Stefano and Modesto, Leonardo and Rachwal, Leslaw and Zhu, Yiwei},
   year={2015},
   month=aug }

@article{Calcagni:2017sov,
    author = "Calcagni, Gianluca and Modesto, Leonardo",
    title = "{Stability of Schwarzschild singularity in non-local gravity}",
    eprint = "1707.01119",
    archivePrefix = "arXiv",
    primaryClass = "gr-qc",
    doi = "10.1016/j.physletb.2017.09.018",
    journal = "Phys. Lett. B",
    volume = "773",
    pages = "596--600",
    year = "2017"
}

@article{Koshelev_2018,
   title={Finite quantum gravity in dS and AdS spacetimes},
   volume={98},
   ISSN={2470-0029},
   url={http://dx.doi.org/10.1103/PhysRevD.98.046007},
   DOI={10.1103/physrevd.98.046007},
   number={4},
   journal={Physical Review D},
   publisher={American Physical Society (APS)},
   author={Koshelev, Alexey S. and Kumar, K. Sravan and Modesto, Leonardo and Rachwał, Lesław},
   year={2018},
   month=aug }

@article{Polyakov:1982ug,
    author = "Polyakov, Alexander m.",
    title = "{PHASE TRANSITIONS AND THE UNIVERSE}",
    doi = "10.1070/PU1982v025n03ABEH004529",
    journal = "Sov. Phys. Usp.",
    volume = "25",
    pages = "187",
    year = "1982"
}

@article{Antoniadis:1991fa,
    author = "Antoniadis, Ignatios and Mottola, Emil",
    title = "{4-D quantum gravity in the conformal sector}",
    reportNumber = "LA-UR-91-1653",
    doi = "10.1103/PhysRevD.45.2013",
    journal = "Phys. Rev. D",
    volume = "45",
    pages = "2013--2025",
    year = "1992"
}

@article{Tsamis:1992sx,
    author = "Tsamis, N. C. and Woodard, R. P.",
    title = "{Relaxing the cosmological constant}",
    reportNumber = "UFIFT-HEP-92-23, CRETE-92-15A",
    doi = "10.1016/0370-2693(93)91162-G",
    journal = "Phys. Lett. B",
    volume = "301",
    pages = "351--357",
    year = "1993"
}

@article{Ford:1984hs,
    author = "Ford, L. H.",
    title = "{Quantum Instability of De Sitter Space-time}",
    reportNumber = "IMPERIAL-TP-83-84-52",
    doi = "10.1103/PhysRevD.31.710",
    journal = "Phys. Rev. D",
    volume = "31",
    pages = "710",
    year = "1985"
}

@article{Mottola:1984ar,
    author = "Mottola, E.",
    title = "{Particle Creation in de Sitter Space}",
    reportNumber = "NSF-ITP-84-123",
    doi = "10.1103/PhysRevD.31.754",
    journal = "Phys. Rev. D",
    volume = "31",
    pages = "754",
    year = "1985"
}

@article{Antoniadis:1985pj,
    author = "Antoniadis, Ignatios and Iliopoulos, J. and Tomaras, T. N.",
    title = "{Quantum Instability of De Sitter Space}",
    reportNumber = "SLAC-PUB-3812",
    doi = "10.1103/PhysRevLett.56.1319",
    journal = "Phys. Rev. Lett.",
    volume = "56",
    pages = "1319",
    year = "1986"
}

@article{Mazur:1986et,
    author = "Mazur, Pawel and Mottola, Emil",
    title = "{Spontaneous Breaking of De Sitter Symmetry by Radiative Effects}",
    reportNumber = "NSF-ITP-85-153",
    doi = "10.1016/0550-3213(86)90058-1",
    journal = "Nucl. Phys. B",
    volume = "278",
    pages = "694--720",
    year = "1986"
}

@article{Abbott:1981ff,
    author = "Abbott, L. F. and Deser, Stanley",
    title = "{Stability of Gravity with a Cosmological Constant}",
    reportNumber = "CERN-TH-3136",
    doi = "10.1016/0550-3213(82)90049-9",
    journal = "Nucl. Phys. B",
    volume = "195",
    pages = "76--96",
    year = "1982"
}

@article{Bruni:2001pc,
    author = "Bruni, Marco and Mena, Filipe C. and Tavakol, Reza K.",
    title = "{Cosmic no hair: Nonlinear asymptotic stability of de Sitter universe}",
    eprint = "gr-qc/0107069",
    archivePrefix = "arXiv",
    doi = "10.1088/0264-9381/19/5/101",
    journal = "Class. Quant. Grav.",
    volume = "19",
    pages = "L23--L29",
    year = "2002"
}

@article{Moreau:2018lmz,
    author = "Moreau, G. and Serreau, J.",
    title = "{Stability of de Sitter spacetime against infrared quantum scalar field fluctuations}",
    eprint = "1808.00338",
    archivePrefix = "arXiv",
    primaryClass = "hep-th",
    doi = "10.1103/PhysRevLett.122.011302",
    journal = "Phys. Rev. Lett.",
    volume = "122",
    number = "1",
    pages = "011302",
    year = "2019"
}

@article{Matsui:2019tlf,
    author = "Matsui, Hiroki and Watamura, Naoki",
    title = "{Quantum Spacetime Instability and Breakdown of Semiclassical Gravity}",
    eprint = "1910.02186",
    archivePrefix = "arXiv",
    primaryClass = "gr-qc",
    reportNumber = "TU-1092",
    doi = "10.1103/PhysRevD.101.025014",
    journal = "Phys. Rev. D",
    volume = "101",
    number = "2",
    pages = "025014",
    year = "2020"
}

@article{Polyakov:2007mm,
    author = "Polyakov, A. M.",
    title = "{De Sitter space and eternity}",
    eprint = "0709.2899",
    archivePrefix = "arXiv",
    primaryClass = "hep-th",
    reportNumber = "PUPT-2244",
    doi = "10.1016/j.nuclphysb.2008.01.002",
    journal = "Nucl. Phys. B",
    volume = "797",
    pages = "199--217",
    year = "2008"
}

@article{Krotov:2010ma,
    author = "Krotov, Dmitry and Polyakov, Alexander M.",
    title = "{Infrared Sensitivity of Unstable Vacua}",
    eprint = "1012.2107",
    archivePrefix = "arXiv",
    primaryClass = "hep-th",
    doi = "10.1016/j.nuclphysb.2011.03.025",
    journal = "Nucl. Phys. B",
    volume = "849",
    pages = "410--432",
    year = "2011"
}
\end{document}